\documentclass[aip,amsmath,amssymb,reprint,graphicx]{revtex4-1}
\usepackage{graphicx}
\usepackage{xcolor}

\draft 

\begin{document}


\title{Theory of Quasi-Statically Screened Electron-Polar Optical Phonon Scattering} 



\author{Yuji Go}
\email[]{yuji.go@warwick.ac.uk}
\author{Rajeev Dutt}
\author{Neophytos Neophytou}
\affiliation{Department of Engineering, University of Warwick}


\date{\today}

\begin{abstract}
The scattering of electrons with polar optical phonons (POP) is an important mechanism that limits electronic transport and determines electron mobility in polar materials. This is typically a stronger mechanism compared to non-polar acoustic and optical phonon scattering, and of similar strength to the Coulomb ionized impurity scattering. At high densities, on the other hand, the cloud of charge carriers screens the dipoles that are responsible for POP scattering, and weakens the electron-POP scattering strength. However, in contrast to ionized impurity scattering, for which the well-known Brooks-Herring equation provides the scattering rates with the effect of screening included, for scattering with POP there is no such closed-form mathematical expression. In this work, we derive such an expression based on Fermi's Golden Rule, which would prove particularly useful in understanding electronic transport in complex crystal and complex band structure materials, in which electron-POP scattering could dominate electronic transport.
\end{abstract}


\maketitle 


\section{Introduction}

The scattering of charge carriers (electrons or holes) with polar optical phonons (POP), is a very important and strong mechanism, seen to have large impact on limiting transport effects in various materials. This well-known mechanism has been used for decades in electronic transport studies. The theory behind it is explained in many instances in the literature since the work of Fröhlich and Lyddane in the late 30's \cite{Frohlich1939-qm,Lyddane_Herzfeld_1938,Ehrenreich,Ridley_2013,nag}. The underlying physics behind POP scattering is linked to the dipole moment caused by the opposite charge on the different atoms in the unit cell of polar materials, which fluctuates due to longitudinal optical (LO) phonon vibrations. Due to its Coulombic nature it is a strong scattering mechanism for electrons, dominating electron transport in typical technologically relevant semiconductors such as GaAs and InAs. More recently, the effect of electron-POP scattering is also the topic of several theoretical first principle calculations as well, which study how the interaction is modified upon doping (typically suppressed as in Ref. \cite{Macheda2022}), compute the dielectric function $\epsilon(\beta,\omega)$ to account for screening \cite{Verdi2017}, or directly compute the interaction Hamiltonian for electron-POP scattering from first principles \cite{Yan2014}.

In addition, the recent advancement in materials synthesis, has allowed the realization of numerous novel materials, most of them polar, targeting a variety of electronic applications. Some examples are the family of Heusler and half-Heusler materials or $\mathrm{Mg_3Sb_2}$ and its derivatives, which are prominent new generation thermoelectric materials \cite{Casper_2012, Quinn_Bos_2021, BERETTA2019100501,Li_Graziosi_Neophytou_2024, VERBRUGGE1993736, Zhang_Song_Iversen_2019, C7MH00865A}. These are polar materials and for many of them, their mobility and electrical conductivity is strongly influenced and limited by electron-POP scattering. In fact, in some cases, such as in $\mathrm{Mg_3Sb_2}$, the POP scattering rates overshadow any other scattering mechanism \cite{Li_Graziosi_Neophytou_2024}.

Since this is a scattering mechanism of Coulombic nature, at high carrier densities it should be mitigated by Coulomb screening effects caused by the cloud of surrounding charge carriers over the dipole charges. The effects of screening in POP are accounted for in electronic transport calculations performed by advanced numerical solvers that employ summations of relevant quantities in k-space, such as the fully ab initio EPW code \cite{epw}, ElecTra \cite{electra}, EPIC STAR \cite{epicstar}, AMSET \cite{amset} and others. 

Typically no matrix elements from ab-initio codes are used for POP directly, as they don’t account for screening. In the codes mentioned above, the Fröhlich interaction term \cite{Frhlich1954,Verdi2015,Sjakste2015} is used for POP scattering. Since these codes deal with the full bandstructure complexity (not just parabolic bands), the Fröhlich equation and the screening terms are numerically applied for all individual k-states (as in this case no analytical closed form expressions are possible).

However, any simplified analytical expressions to describe the POP scattering rates as a function of energy (and for single parabolic bands), do not yet include the effects of screening. Thus, in general they overestimate the scattering rates and underestimate the material mobility and conductivity at high densities. Despite the fact that for ionized impurity scattering (IIS) the screening effect is well accounted for by the model proposed by Brooks and Herring long ago \cite{brooks_herring}, there are no instances of a similarly closed-form analytical expression for POP scattering that includes screening.

This work aims to provide such an expression, that can accurately capture the effect of screening in electron-POP scattering for a single parabolic band, in the same way that the Brooks-Herring model describes screened IIS scattering. Starting from Fermi's Golden Rule we derive an equation that includes the generalized screening term \cite{Ridley_2013,nag,Yu_Cardona_2010} and analyze how it behaves under different conditions, such as low and high densities and temperatures, as well as how it compared to the IIS scattering rates. We further present the characteristic mobility exponents under POP scattering, which are essential in evaluating the internal transport properties of materials in experimental works. 

Our final expression is more complicated in comparison to the Brooks-Herring formula due to the inelastic and anisotropic nature of POP, which includes both emission and absorption of phonons. However, it can still be used in closed form to describe POP scattering rates and electronic transport in polar materials more accurately. 

The paper is organized as follows: In Section II we present the electron-POP scattering theory and derive the formula for screened scattering rates. In Section III we provide results for the trends described by our derived formula with regards to carrier density, temperature, the characteristic exponents that describe the mobility trends, and comparisons to the unscreened POP case and the well-established IIS formula for charged impurity scattering. In Section IV we provide an elaborative discussion on the regions of validity of the formalism derived. Finally, in Section V we summarize and conclude.

\section{Theory}

\subsection{Theory of unscreened POP scattering}

We start our derivation by following the standard unscreened POP scattering rate derivation as presented by Lundstrom in Ref. \cite{lundstrom} and elsewhere for completeness, and then derive the screened rate expression. 

\begin{figure*}
    \centering
    \includegraphics[width=\textwidth]{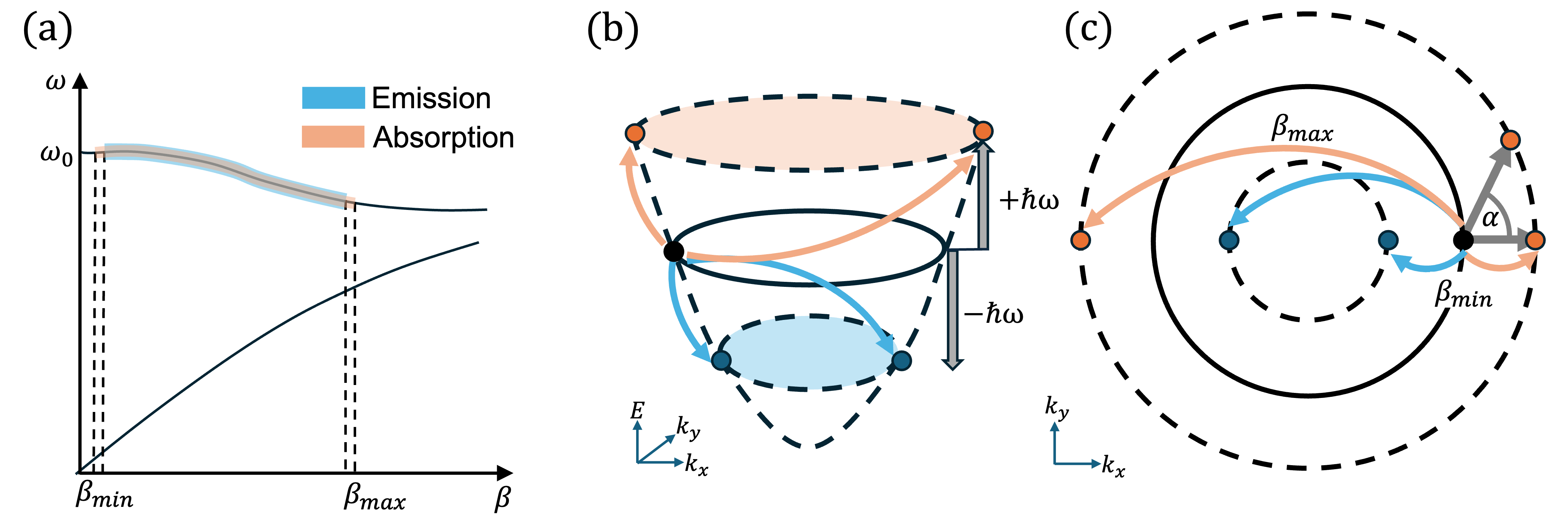}
    \caption{(a) Phonon spectrum with indication for the wavevector limits $\beta_{max}$ and $\beta_{min}$ that dictate the inelastic optical phonon scattering process. (b) 3D schematic for the electronic band, illustrating the emission and absorption scattering process from an initial state (middle part).(c) 2D cross section schematic for the optical phonon scattering process, indicating the the wavevector limits $\beta_{max}$ and $\beta_{min}$ and the scattering angle $\alpha$ for a given (arbitrary) scattering event.}
    \label{fig:fig1}
\end{figure*}

During a scattering event, a charge carrier with an initial state of energy $E$ and momentum $p$, scatters into a new (final) state of energy $E'$ and momentum $p'$. POP is an inelastic, anisotropic mechanism. The scattering event is illustrated using the change in energy $\Delta E$ (absorption and emission with $\hbar\omega$ energy change), the polar scattering angle $\alpha$, as well as the exchange vector $\beta$, as shown in Figure \ref{fig:fig1} (b) and (c). Note that this process happens in 3D momentum space, whereas in Figure 1 this is illustrated for a 2D momentum space for clarity. It is typical in the literature to use the symbol $q$ for the phonon exchange vector, but here we use $\beta$ as in Ref. \cite{lundstrom}, with which we attempt to establish continuity.

The associated transition rate from state $p$ to $p'$ is governed by Fermi's Golden rule as:  

\begin{equation}
    S(p,p') = \frac{2\pi}{\hbar}  |H_{p',p}|^2 \delta(E(p') - E(p) - \Delta E)
    \label{equation:fgr}
\end{equation}

 where $H_{p',p}$ is the matrix element that describes the effect of an external perturbing potential $U$ on the band states. The potential $U$ is obtained using the deformation potential approximation, where phonons cause deformations in the lattice structure, leading to distortions of the local conduction/valence band edge. The deformation is described by an one-dimensional elastic wave $u(x,t)$ as:

\begin{equation}
    u(x,t) = A_\beta e^{i(\beta x-\omega t)} + A_\beta^* e^{i(-\beta x-\omega t)}
    \label{equation:u_el}
\end{equation}

Electron - Polar Optical Phonon (POP) scattering occurs in materials with two or more different atoms in the unit cell, creating a dipole moment with effective charge $q^*$. This charge in general differs from the true ionic charge due to the distortion of the electronic shells and the electric fields that are set up \cite{Callen_1949}. When the atoms are deformed via longitudinal phonon vibrations, the dipole moment varies as well, creating an accumulation of positive and negative ions, leading to a Coulomb potential $U$ and electric fields within the crystal, which can interact with and scatter electrons. The potential $U$ is absent for transverse phonons, as the direction of polarization is perpendicular to the displacement caused by these phonons. In addition, only the longitudinal optical modes are effective in scattering electrons, while the longitudinal acoustic ones have a much weaker effect (piezoelectric effect). Thus, we only consider longitudinal optical phonons, for which the displacement $u(x,t)$ perturbs the dipole moment $p_d$ directly proportional to its amplitude as:

\begin{equation}
    \delta p_d = q^* u
    \label{equation:delta_p}
\end{equation}

The dipole electric field $\mathcal{E}$ is related to the dipole moment via:  

\begin{equation}
    D = \epsilon_0 \mathcal{E} + \delta P
    \label{equation:e_disp}
\end{equation}

where $D$ is the electric displacement field and $\delta P$ is the dipole moment charge per unit volume ($\delta p_d/V_u$) where $V_u$ is the volume of the unit cell. Using the spatial displacement $u$ from Eq. \ref{equation:u_el}, the displacement field $D$ is given by:

\begin{equation}
    D(x) = De^{i\beta x} + D^*e^{-i\beta x} = |D|cos(\beta x+ \phi)
\end{equation}

By assuming there are no resultant free charges (i.e. $\nabla \cdot D = 0$), we find that $D = 0$. Thus, the electric field can be written in the form of:

\begin{equation}
    \mathcal{E} = - \frac{\delta P}{\epsilon_0} = -\frac{q^* u}{\epsilon_0 V_u}
    \label{equation:e_2}
\end{equation}

By integrating Eq. \ref{equation:e_2} over all space after substituting the $u$ term from the elastic wave described in Eq. \ref{equation:u_el}, the external perturbing potential for POP, $U_{POP}$, is obtained as:

\begin{equation}
    U_{POP} = -q\int \mathcal{E} dx = \frac{qq^*u}{i\beta V_u \epsilon_0}
    \label{equation:pop_u}
\end{equation}

It is now useful to examine the value of the effective charge and its dependencies. The effective charge, $q^*$, can be obtained by examining the difference in energy caused by the longitudinal and transverse modes. The derivation below follows the work of Callen in Ref \cite{Callen_1949}. The longitudinal and transverse modes are considered as a simple oscillating system. In the absence of dipole moments (i.e. in a non-polar material), the two have identical energies (i.e. the longitudinal and transverse modes are degenerate). In general, the total energy density, $U_E$, of an oscillating system can be classically expressed as: 

\begin{equation}
    U_E = \frac{1}{2}\frac{M\omega^2}{V}u^2
\end{equation}

where $M$ is the reduced mass of the ions, $\omega$ is the frequency of the system, and $V$ is the volume of the unit cell. In the presence of dipole moments (i.e. in a polar material), starting from Eq. \ref{equation:e_2}, the electrostatic energy density, $U_q$, associated with the produced electric field is given by: 

\begin{equation}
    U_q = \frac{1}{2}\epsilon_0\kappa_\infty |\mathcal{E}|^2 = \frac{1}{2}\frac{\kappa_\infty q^{*2}u^2}{V^2\epsilon_0}
\end{equation}

In the presence of dipole moments in polar materials, this electrostatics energy is added to the longitudinal mode $U_L$ (only). This is because the longitudinal mode experiences an additional electric field, while the transverse mode, $U_T$, does not because the polarization is perpendicular to the direction of propagation of transverse waves \cite{Frohlich1939-qm,Lyddane_Herzfeld_1938}. Equating the energy densities of the two modes together ($U_L = U_T + U_q$), we obtain:

\begin{equation}
    \frac{1}{2}\frac{M\omega_L^2}{V}u^2 = \frac{1}{2}\frac{M\omega_T^2}{V}u^2 + \frac{1}{2}\frac{\kappa_\infty q^{*2}}{V^2\epsilon_0}u^2
    \label{equation:e_density_sum}
\end{equation}

\begin{equation}
    \omega_L^2 = \omega_T^2 + \frac{\kappa_\infty q^{*2}}{MV\epsilon_0}
    \label{equation:omega_L}
\end{equation}

where $\omega_L$ and $\omega_T$ are the characteristic frequencies of the respective modes. 

If an external field $D$ is applied in parallel to the displacement $u$, it will induce an additional electronic polarization $P$ and alter the electric field $\mathcal{E}$ (from the form of Eq. \ref{equation:e_disp}) as:

\begin{equation}
    P = \frac{q^*u}{V\epsilon_0} +\bigg(\frac{\kappa_{\infty}-1}{\kappa_\infty}\bigg)D
    \label{equation:p_new}
\end{equation}
\begin{equation}
    \mathcal{E} = -\frac{q^*u}{V\epsilon_0} + \frac{D}{\epsilon_0\kappa_\infty}
\end{equation}

where $\kappa_\infty$ is the electronic component of the dielectric constant. The total energy density would then increase to:
\begin{equation}
    U = \frac{1}{2}\frac{M}{V}\omega_T^2u^2 + \frac{1}{2}\epsilon_0\kappa_\infty\bigg[\frac{-q^*u}{V\epsilon_0}+\frac{D}{\epsilon_0\kappa_\infty}\bigg]^2
\end{equation}
Under the external field $D$, the ions would rearrange themselves to achieve minimum energy density. Taking the steady state solution ($\frac{dU}{du}=0$), one finds that: 

\begin{equation}
    u = \frac{q^*V}{MV\omega_T^2\epsilon_0+q^{*2}\kappa_\infty}D
    \label{equation:disp_new}
\end{equation}

Substituting Eq. \ref{equation:disp_new} into Eq. \ref{equation:p_new}, the total polarization is given by:

\begin{equation}
    P = \frac{q^{*2}}{MV\omega_T^2\epsilon_0^2+q^{*2}\epsilon_0\kappa_\infty}D + (\frac{\kappa_{\infty}-1}{\kappa_\infty})D
\end{equation}

The total polarization can also be expressed in the form ($P = \frac{\kappa_0-1}{\kappa_0}D$), where $\kappa_0$ is the static dielectric constant, accounting for both the electronic and ionic contributions (an overall effective value to account for all contributions). Substituting for $\omega_T$ from Eq. \ref{equation:omega_L} and rearranging, we obtain an expression for the effective charge $q^*$ as:

\begin{equation}
    q^* = \sqrt{\frac{\epsilon_0VM\omega_{L}^2}{\kappa_0}(\frac{\kappa_0}{\kappa_\infty}-1)}
    \label{equation:qstar}
\end{equation}

Further, substituting Eq. \ref{equation:qstar} into Eq. \ref{equation:omega_L} gives us the transcendental equation \cite{Callen_1949,Ehrenreich}

\begin{equation}
    \omega_L^2 = \omega_T^2 \frac{\kappa_0}{\kappa_\infty}
\end{equation}
 
To quantify the effective charge, for instance, NbFeSb, a prominent Half-Heusler has $q^* = 0.39eV$, while GaAs, a material known for its strong polar nature, has $q^* = 0.51eV$ \cite{Callen_1949}. Values for other materials as well are listed in Table \ref{table:ef_charge} of Appendix \ref{chapter:eff_charge}. From here onwards, the longitudinal frequency $\omega_L$ is expressed as $\omega_0$ following Lundstrom \cite{lundstrom}.

The external potential is then used to form the scattering matrix element as:

\begin{equation}
    H_{p',p} = \frac{1}{\Omega} \int^{\infty}_{\infty} e^{-i p' \cdot r / \hbar} U(r) e^{i p\cdot r / \hbar} dr^3
    \label{equation:h_integral}
\end{equation}

The potential can be written in the form $U = K_\beta u_\beta=K_\beta A_\beta e^{\pm i (\beta\cdot r - \omega_0 t)}$, which allows us to write the matrix element in the form:

\begin{equation}
    |H_{p',p}|^2 = |K_\beta|^2|A_\beta|^2 \delta_{p',p+\hbar\beta}
    \label{equation:h_integral_2}
\end{equation}

where $|A_\beta|$ is the magnitude of the lattice vibration (see below), and $K_\beta$ is a mechanism-specific dependent term derived from the external potential $U$. As described in Ref. \cite{lundstrom}, $K_\beta$ for unscreened POP scattering is given by

\begin{equation}
    |K_\beta|^2 = \frac{q^2q^{*2}}{V^2\beta^2\epsilon_0^2}=\frac{M q^2 \omega_0^2}{V\epsilon_0 \kappa_0\beta^2}(\frac{\kappa_0}{\kappa_\infty} -1)
    \label{equation:k_beta}
\end{equation}
 
 $|A_\beta|$ can be expressed using a quantum mechanical approach \cite{Datta_1989}. Starting with the elastic wave (Eq. \ref{equation:u_el}), one can also express $u(x,t)$ as:

\begin{eqnarray}
    u(x,t) = A_\beta e^{i(\beta x-\omega_0 t)} + A_\beta^* e^{i(-\beta x-\omega_0 t)} \nonumber \\ =  2|A_\beta|cos(\beta x-\omega_0 t + \phi)
\end{eqnarray}

The maximum mechanical energy of the system is given as:

\begin{equation}
    E_{K} = \frac{1}{2}M|\frac{du}{dt}|^2 = \frac{1}{2}\frac{M}{V}\Omega\omega_0^24|A_\beta|^2
    \label{equation:ke_max}
\end{equation}

Similarly, the system can be treated as a quantum oscillator, where the energy is given by $E = (N+1/2)\hbar\omega_0$. Thus, $|A_\beta|$ can be given by:

\begin{equation}
    |A_\beta|^2 \approx \frac{\hbar V}{2 M\omega_0} (N_\omega + \frac{1}{2})
    \label{equation:a_beta}
\end{equation}

where $\Omega$ is the volume of the crystal, and $N_\beta$ is phonon occupation number given the Bose-Einstein phonon distribution function. However, it is known that using the term $(N_\omega + \frac{1}{2})$ is inaccurate, and separate treatments for the cases of emission and absorption are needed, as:
\begin{equation}
    |A_\beta|^2 \approx \frac{\hbar V}{2M\Omega\omega_0} \Bigl(N_\omega + \frac{1}{2} \mp \frac{1}{2}\Bigl)
    \label{equation:a_beta_2}
\end{equation}

Hence, the matrix element is given by: 

\begin{equation}
    H_{p',p} = \frac{qq^*}{\epsilon_0\beta}\sqrt{\frac{\hbar(N_\omega+\frac{1}{2} \mp \frac{1}{2})}{2VM\Omega\omega_0}}
    \label{equation:h_pop}
\end{equation}

Further manipulation of the Kronecker delta in Eq. \ref{equation:fgr} provides the final expression for the transition rate as:

\begin{equation}
    S(p,p') = C_\beta \Bigl(N_\omega + \frac{1}{2} \mp \frac{1}{2} \Bigl)\delta\Bigl(\pm cos(\theta) + \frac{\hbar}{2p} \mp \frac{\omega_0}{\nu\beta}\Bigl)
    \label{equation:s_2}
\end{equation}

where $C_\beta$ is a mechanism-specific variable, which combines $|K_\beta|^2$ alongside additional terms from $|A_\beta|^2$ and the modified Kronecker delta. The $C_\beta$ for unscreened polar optical phonon (POP) scattering is given by:

\begin{equation}
    C_{\beta} = \frac{\pi m^* q^2 \omega_0}{\hbar \kappa_0 \epsilon_0 \beta^3 p \Omega}\Bigl(\frac{\kappa_0}{\kappa_\infty} -1\Bigl)
    \label{equation:pop_nos}
\end{equation}
 
The total scattering rate is given by summing $S(p,p')$ over all possible final scattering states $p'$ as:

\begin{equation}
    \frac{1}{\tau(E)} = \sum_{p'}S(p,p') = \sum_{\beta}S(p,p')
\end{equation}

The second part of the equation is the notation for the case of a parabolic band, as assumed in this work. Here $\beta$ is the phonon wavevector since the mapping from $p'$ to $\beta$ is unique \cite{lundstrom}. 

Upon a scattering event, both energy and momentum conservation need to be satisfied between the initial and final states and the participating phonon. Here we assume a single optical phonon of energy value $\omega_0$. As illustrated in Figure \ref{fig:fig1}, a polar optical phonon scattering event results in phonon emission or phonon absorption by the electron. A finite phonon energy imposes a finite momentum $\beta$ upon a transition from an initial to a final scattering state. This imposes limits on maximum and minimum allowed phonon vectors, $\beta$, for the scattering events at a particular energy, $E$, as dictated by the geometrically closest and farthest possible final states from the initial state in k-space. As seen in Figs. \ref{fig:fig1}(a-c), there is a minimum and maximum $\beta$ value, which is also different for the emission and absorption processes. The values can be extracted from the simple spherical geometry of a parabolic, isotropic band upon substitution of the $E(k)$ relation, and are given, respectively, by:

\begin{equation}
    \beta_{max} = \frac{p}{\hbar}\biggl(1 + \sqrt{1 \pm \frac{\hbar\omega_0}{E}}\biggl)
    \label{equation:beta_max}
\end{equation}

\begin{equation}
    \beta_{min} = \frac{p}{\hbar}\biggl(\mp1 \pm \sqrt{1 \pm \frac{\hbar\omega_0}{E}}\biggl)
    \label{equation:beta_min}
\end{equation}

What determines electronic transport is not the relaxation time itself, but the momentum relaxation time $\tau_m$, given by:

\begin{equation}
    \frac{1}{\tau_m(E)} = \sum_{p'}S(p,p')\Bigr[1 - \frac{p'}{p}cos(\alpha)\Bigr]
    \label{equation:momentum_relaxation}
\end{equation}

where $\alpha$ is the polar angle between the incident and scattered states.

After inserting Eq. \ref{equation:s_2} into Eq. \ref{equation:momentum_relaxation} and performing standard simplifications, the momentum relaxation time can be evaluated by

\begin{equation}
    \frac{1}{\tau_m(E)} = \frac{\Omega}{4\pi^2} \int^{\beta_{max}}_{\beta_{min}} \Bigl(N_{\omega} + \frac{1}{2} \mp \frac{1}{2}\Bigl)\Bigl( C_{\beta} \Bigl)\Bigl(\frac{\hbar \beta}{2p} \mp \frac{\omega_0}{\nu \beta}\Bigl) \frac{\hbar \beta^3}{p} d\beta
    \label{equation:tau_m}
\end{equation}

where $C_\beta$ is the mechanism specific variable given in Eq. \ref{equation:pop_nos}, $p$ and $v$ are the momentum and velocity of the electron. The resultant scattering rate for unscreened POP is given by (the standard expression in the literature \cite{lundstrom}) 
\begin{widetext}
    \begin{eqnarray}
    \frac{1}{\tau_m(E)} = \frac{m^*q^2\omega_0}{4\pi\kappa_0 \epsilon_0 p^2}\Bigl(\frac{\kappa_0}{\kappa_\infty}-1\Bigl)\Bigl(N_\omega + \frac{1}{2} \mp \frac{1}{2}\Bigl)\biggr[\frac{\hbar}{4p}(\beta_{max}^2 - \beta_{min}^2) \mp\frac{\omega_0}{\nu} ln\Bigl(\frac{\beta_{max}}{\beta_{min}}\Bigl)\biggr]
    \label{equation:pop_ns_full}
    \end{eqnarray}
\end{widetext}

\subsection{Theory of screened POP scattering}

Nag \cite{nag} describes the effect of Coulomb screening on scattering potentials. It was shown that all scattering potentials/matrix elements can be scaled by a factor $\beta^2/(\beta^2 + 1/L_{TF}^2)$, which is referred to as a generalized screening term. We provide a derivation of this term by partially following the original work by Ehrenreich \cite{Ehrenreich} and a more recent one by Ridley \cite{Ridley_2013} which involves Lindhard function theory \cite{Ehrenreich1959}, in Appendix \ref{chapter:screening_term}. This treatment is valid when the thermal velocity of the electrons is larger than the velocity of lattice waves, and can redistribute themselves in reaction to the varying potential. Essentially, to be able to screen, the electron gas needs to be able to respond fast compared to the oscillation of the perturbing potential (this is determined by the phonon frequency). We elaborate in detail on the validity of this approach further below in Section IV. Above, $L_{TF}$ is the Thomas-Fermi screening length, describing the screening distance of the decaying potential. The expression for $L_{TF}$ is given by \cite{lundstrom,NeophytouPRBNWs}:

\begin{equation}
    L_{TF} =  \sqrt{\frac{\kappa\epsilon_0 k_B T}{q^2n}\frac{\mathcal{F}_{1/2}(\eta)}{\mathcal{F}_{-1/2}(\eta)}}
    \label{equation:l_tf}
\end{equation}

where $\kappa$ is the relative dielectric constant, $n$ is the charge density, and $\mathcal{F}_{1/2}(\eta)$ and $\mathcal{F}_{-1/2}(\eta)$ are Fermi-Dirac integrals of order $1/2$ and $-1/2$ respectively with $\eta = \frac{\eta_F}{k_BT}$ where $\eta_F$ is the reduced Fermi level ($\eta_F = E_F - E_C$). The derivation for $L_{TF}$ is given in Appendix \ref{chapter:l_tf}. In $L_{TF}$, for POP scattering we use the high-frequency dielectric constant, i.e. $\kappa=\kappa_\infty$, as the screening is caused by the electron cloud, while for IIS scattering we use the static dielectric constant, i.e. $\kappa=\kappa_0$. Note that here we use the Thomas-Fermi screening length as it is valid in both the non-degenerate and degenerate regimes. In the non-degenerate regime, the Debye screening length approximation is typically used, which can be expressed in a more simple manner as a function of the density alone. As we show in the Appendix \ref{chapter:l_tf}, the two methods provide similar results under non-degenerate conditions, but deviate for higher densities. Here, the Fermi-Dirac integrals were computed using a Python implementation of the algorithm created by Mohan et. al \cite{mohan95}. The screened matrix element $H(p',p)$ is then given by:

\begin{equation}
    H_{p',p} = \frac{qq^*}{\epsilon_0}\sqrt{\frac{\hbar(N_\omega+\frac{1}{2} \mp \frac{1}{2})}{2VM\Omega\omega_0}} \frac{\beta}{\beta^2+\frac{1}{L_{TF}^2}}
    \label{equation:h_pop_s}
\end{equation}

\begin{figure}[h]
    \centering
    \includegraphics[scale=0.15]{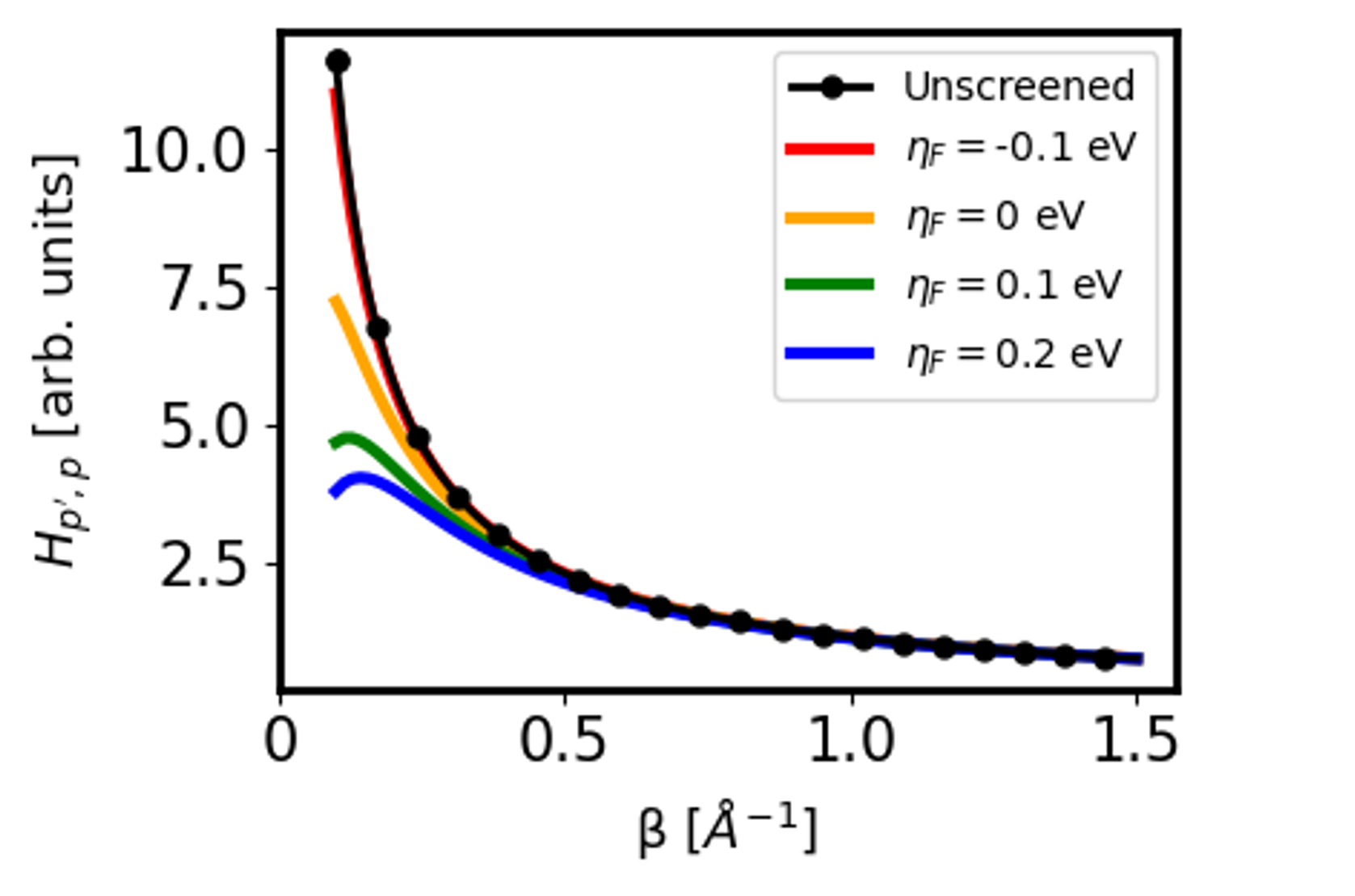}
    \caption{The matrix elements for the unscreened and screened POP scattering for different $\eta_\mathrm{F}$ values. Modeled by a material with $\kappa_0 = 42.83$, $\kappa_\infty = 23.91$, $\hbar\omega_0 = 36$ $meV$.}
    \label{fig:fig2}
\end{figure}

The matrix element trend is shown in Fig. \ref{fig:fig2} versus $\beta$, for different values of $\eta_F$. In the unscreened case (or low $\eta_F$ case), the matrix element has the well-known divergence for small $\beta$. As the $\eta_F$ increases and screening comes into effect, the matrix elements reduce. For lower $\beta$ values, the singularity is now gradually eliminated and the matrix elements tend towards zero for zero $\beta$. Note that we present this in arbitrary units, since the matrix element is a function of the unit cell volume $\Omega$, which is arbitrary in our formalism.

To determine the scattering rates now, the $C_\beta$ term for screened POP scattering is given by:

\begin{equation}
    C_{\beta} = \frac{\pi m^* q^2 \omega_0}{\hbar \kappa_0 \epsilon_0 \beta^3 p \Omega}\Bigl(\frac{\kappa_0}{\kappa_\infty} -1\Bigl) \frac{\beta^4}{(\beta^2 + \frac{1}{L_{TF}^2})^2} \text{\hspace{10pt} (POP screened)}
    \label{equation:pop_s_beta}
\end{equation}

which is the same as Eq. \ref{equation:pop_nos}, but with the additional generalized screening term squared included. 

Substituting $C_\beta$ into Eq. \ref{equation:tau_m}, we obtain: 

\begin{widetext}
    \begin{eqnarray}
    \frac{1}{\tau_m(E)}=\frac{m^* q^2 \omega_0}{4 \pi \kappa_0 \epsilon_0 p^2}\Bigl(\frac{\kappa_0}{\kappa_\infty}-1\Bigl) \Bigl(N_\omega + \frac{1}{2} \mp \frac{1}{2}\Bigl)\int^{\beta_{max}}_{\beta_{min}} \frac{1}{(\beta^2+\frac{1}{L_{TF}^2})^2} \Bigl(\frac{\hbar\beta^5}{2p} \mp \frac{\omega_0 \beta^3}{\nu} \Bigl) d\beta
    \label{equation:pops_1}
    \end{eqnarray}
\end{widetext}

where $\beta_{min}$ and $\beta_{max}$ are given by Eq. \ref{equation:beta_min} and \ref{equation:beta_max} respectively. The above expression requires the evaluation of two integrals, namely:

\begin{equation}
    I_1 = \int \frac{\beta^5}{(\beta^2+\frac{1}{L_{TF}^2})^2}d\beta
    \label{equation:int1}
\end{equation}

\begin{equation}
    I_2 = \int \frac{\beta^3}{(\beta^2+\frac{1}{L_{TF}^2})^2}d\beta
    \label{equation:int2}
\end{equation}

The solution for Eq. \ref{equation:int2} can be found in integral tables, however Eq. \ref{equation:int1} must be solved manually. The details of the integral evaluations are given in Appendix \ref{chapter:integrals}, while the final solutions are given, respectively, as:

\begin{equation}
    I_1 = \Bigr[ \frac{-\beta^4}{2(\beta^2 + \frac{1}{L_{TF}^2})} + \frac{1}{L_{TF}^2}(1+\beta^2L_{TF}^2 -ln|1+L_{TF}^2\beta^2|)\Bigr]^{\beta_{max}}_{\beta_{min}}
    \label{equation:int1_sol}
\end{equation}

\begin{equation}
    I_2 = \Bigr[ \frac{-1}{2}\frac{\beta^2L_{TF}^2}{1+\beta^2L_{TF}^2} +\frac{1}{2}ln|1 + L_{TF}^2\beta^2|\Bigr]^{\beta_{max}}_{\beta_{min}}
    \label{equation:int2_sol}
\end{equation}

Substituting these solutions into Eq. \ref{equation:pops_1} above and expanding, we obtain Eq. \ref{equation:pop_s_full} below as the final expression for the screened POP momentum relaxation scattering rates:

\begin{widetext}
    \begin{eqnarray} 
        \frac{1}{\tau_m} = \frac{m^*q^2\omega_0}{4\pi\kappa_0 \epsilon_0 p^2}\Bigl(\frac{\kappa_0}{\kappa_\infty}-1\Bigl)\Bigl(N_\omega + \frac{1}{2} \mp \frac{1}{2}\Bigl)\biggr[\frac{-\hbar}{4p}(\frac{\beta_{max}^4}{\beta_{max}^2 + \frac{1}{L_{TF}^2}} - \frac{\beta_{min}^4}{\beta_{min}^2 + \frac{1}{L_{TF}^2}}) + \frac{\hbar}{2p}(\beta_{max}^2 - \beta_{min}^2)  \nonumber \\
        + \Bigl(\frac{-\hbar}{2pL_{TF}^2} \mp \frac{\omega_0}{2v}\Bigl)ln\Bigl(\frac{\beta_{max}^2 + \frac{1}{L_{TF}^2}}{\beta_{min}^2 + \frac{1}{L_{TF}^2}}\Bigl) \pm \frac{\omega_0}{2v}\Bigl(\frac{\beta_{max}^2}{\beta_{max}^2 + \frac{1}{L_{TF}^2}} - \frac{\beta_{min}^2}{\beta_{min}^2 + \frac{1}{L_{TF}^2}}\Bigl)\biggr] 
        \label{equation:pop_s_full}
    \end{eqnarray}
\end{widetext}

where again $\beta_{max}$ and $\beta_{min}$ are given by Eq. \ref{equation:beta_max} and \ref{equation:beta_min}, respectively.

\section{Discussion}

\subsection{Screened POP scattering rate trends}

\begin{figure*}
    \centering
    \includegraphics[width=\textwidth]{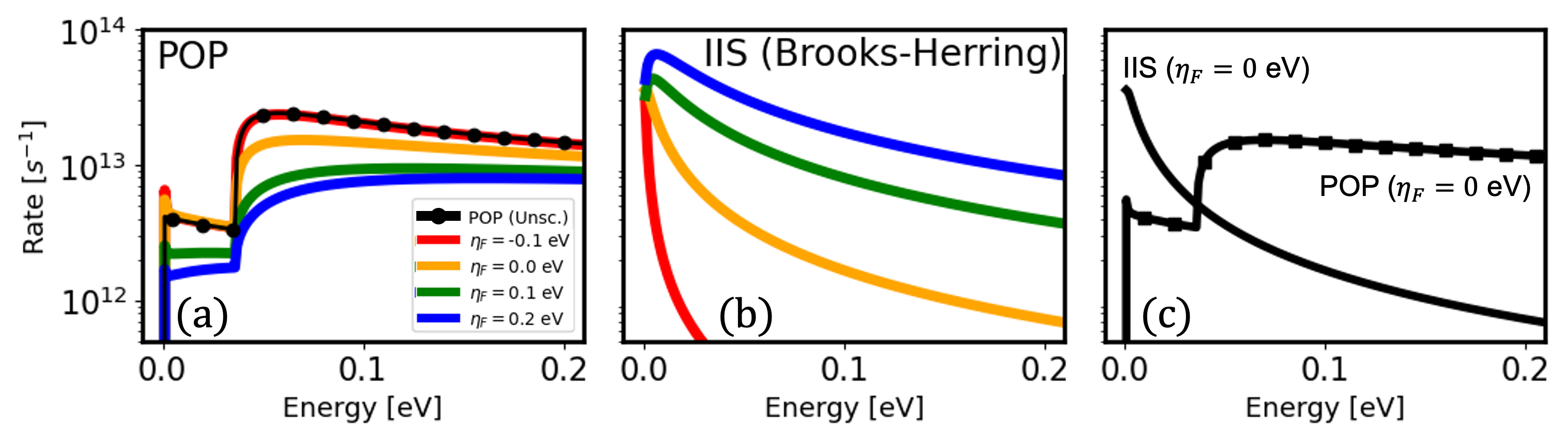}
    \caption{(a) Scattering rates for POP scattering. The solid colored lines show POP with screening (with the Fermi Energy position indicated in the legend). The black solid-dot line shows POP without screening. (b) Scattering rates for IIS (Brooks-Herring model) for the same Fermi level positions as in (a). (c) Comparison between POP and IIS scattering rates at $\eta_F = 0 \mathrm{eV}$.}
    \label{fig:fig3}
\end{figure*}

The screened POP scattering rates are plotted versus energy, $E$, for various reduced Fermi level values $\eta_F$, in Figure \ref{fig:fig3}. For the purpose of this work, a polar material, modeled by a parabolic band with $m^*=0.92 m_0$, $\hbar\omega_0=36 meV$ and with dielectric constants $\kappa_0=42.83$ and $\kappa_\infty=23.91$ was used and simulated (without loss in generality - but subject to the regions of validity depending on the Fermi level, temperature, and material specific parameters as discussed below) \cite{material_descriptors}. This corresponds to the half-Heusler NbFeSb, which is a prominent thermoelectric material \cite{nbfesb1,nbfesb2,nbfesb3}. The solid lines in Figure \ref{fig:fig3}(a) show the screened POP scattering rates as derived above, while the black-dotted line plots the unscreened POP scattering rate. The rates start with phonon absorption at lower energies, whereas at the onset of phonon emission a discontinuity is observed in the rates, signifying the additional process, which is also a stronger one due to the inclusion of zero point fluctuations. The screened and unscreened solutions coincide at low $\eta_F$, which is the limit of the unscreened case, indicating that our derived formula behaves as expected. As $\eta_F$ increases, i.e. as the carrier density in the material increases, the screening becomes stronger, decreasing the scattering rate. At high $\eta_F$ the system enters the degenerate regime and requires the consideration of Fermi-Dirac statistics in $L_{TF}$. As seen in Figure \ref{fig:figS3} in Appendix \ref{chapter:l_tf}, in the degenerate regime the decrease in $L_{TF}$ is seen to stagnate, leading to the scattering rates tending towards saturation at high $\eta_F$.

\subsection{Comparisons to Ionized Impurity Scattering (IIS)}

It is illustrative to compare the screened POP rates that we have reached, to the well established screened Ionized Impurity Scattering (IIS), as provided by the Brooks-Herring formula \cite{brooks_herring}. IIS is another strong scattering mechanism caused by electrostatic interactions, which occurs when an electron is scattered by the electric field of an ionized impurity. The electric field from the ionized impurity can be screened by the surrounding charge carriers. In this case, the perturbing potential is given by a combination of a Coulombic potential and an exponential term dependent on the Thomas-Fermi (screening) length, $L_{TF}$, as:

\begin{equation}
    U_s(r) = \frac{q^2}{4\pi\kappa_0\epsilon_0r} e^{-r/L_{TF}}
    \label{equation:iis_u}
\end{equation}

The matrix element $H_{p',p}$ is given by:

\begin{equation}
    H_{p',p} = \frac{q^2}{\Omega\kappa_0\epsilon_0}\frac{1}{\beta^2+\frac{1}{L_{TF}^2}}
    \label{equation:h_iis}
\end{equation}

In this case, the IIS momentum relaxation time $\tau_m(E)$ is given by (as per standard literature result \cite{lundstrom}):

\begin{equation}
    \tau_m(E) = \frac{16\sqrt{m^*}\pi\kappa_0^2\epsilon_0^2}{N_1q^4}\Bigr[ln(1+\gamma^2) - \frac{\gamma^2}{1+\gamma^2}\Bigr]^{-1} E^{3/2}
    \label{equation:iis_tau}
\end{equation}

\begin{equation}
    \gamma = \frac{8m^*L_{TF}^2E}{\hbar^2}
    \label{equation:gamma}
\end{equation}

Figure \ref{fig:fig3}(b) plots the IIS rates. At low carrier densities (negative $\eta_F$) the IIS rates decrease with energy, with the first two $\eta_F$ we use (red and orange lines), indicating the well-known singularity of IIS at low energies. For higher densities, (high $\eta_F$, green and blue lines), the rates increase globally, while the singularity is now removed, with the rates acquiring a downward trend near $E=0$ eV. In this regime, screening would dominate the external potential, transforming it towards a delta-function, and in this case it is expected that the rate tend towards an energy dependence of $\tau^{-1} \propto E^{1/2}$, typical of the density of states dependence. This is analogous to the case of strong Coulomb screening, as discussed later \cite{lundstrom}. In our simulations, however, the overall trend is found to be negative regardless of charge density $\eta_F$, suggesting the model (and densities considered) does not enter the strong screening regime. We might be seeing the beginning of that regime at very low energies fo the larger $\eta_F$ values, but in general, the overall trend is downward (although it becomes shallower with energy).

As $\eta_F$ increases, screening increases, which tends to reduce the scattering rates. This is the case for screened POP scattering, where the rates monotonically decrease across energies with increasing charge density, as shown in Figure \ref{fig:fig3}(a). For IIS, however, the scattering rates are also directly proportional to the density of ionized impurities, which is equal to the carrier density in typical cases and increases the rate back. In the example we consider, the latter is dominant and IIS increases across energies with increasing density. Another observation is the much larger spread of the IIS rates with $\eta_F$ compared to the POP spread. The spread in IIS rates between the low and high densities we consider is around two orders of magnitude, wheres for the same densities in POP is less than one order of magnitude.  

It is also interesting to compare quantitatively the POP and IIS rates as shown in Figure \ref{fig:fig3}(a) and \ref{fig:fig3}(b), respectively. For high densities, the IIS rates are higher compared to the POP rates in most of the energy range considered. For intermediate and lower densities (and $\eta_F$), the comparison is not straightforward. Figure \ref{fig:fig3}(c) compares the two rates at ${\eta_F = 0.0eV}$, a typical condition for highly doped semiconductors (and where the thermoelectric power factor peaks, for example). Here, at low energies IIS is stronger due to the rates tending towards a singularity, but at higher energies IIS drops fast and POP takes the lead (note though that we consider a strongly polar material).

In general, POP seems to either drop less sharply (at low $\eta_F$), or increase more (at higher $\eta_F$) with energy, compared to IIS, which drops much faster in the majority of the energy range. Comparing the matrix elements for screened POP (Eq. \ref{equation:h_pop_s}) and IIS (Eq. \ref{equation:h_iis}), the POP matrix element has an additional $\beta$ dependence. At high energy the $\beta$ values are larger, leading to larger matrix elements for POP compared to IIS. This results in larger increasing trends (or slowing decreasing trends) for POP rates compared to IIS in general. Note that POP and IIS depend on specific independent parameters, thus quantitative comparisons of the scattering rates might not be fully meaningful, but our comparisons provide a first order indication of the qualitative differences in trends, and how the POP scattering rates in a strongly polar material compare to the IIS rates.

In addition to the Brooks-Herring formula, in the limits of weak ($L_{TF} \to \infty$) and strong screening ($L_{TF} \to 0$), IIS is described by different standard literature expressions. The weak screening limit, called the Conwell-Weisskopf approximation \cite{Conwell_Weisskopf_1950}, is derived by approximating the electron to be a localized wave packet that moves with a classical orbit and scattering is described by a geometrical impact parameter \cite{Ridley_2013}. By following the laws of classical scattering theory, the model avoids the divergence as observed in Figure \ref{fig:fig3}(b) for low densities. The equation for the CW model is given by:

\begin{equation}
    \frac{1}{\tau_m} = \frac{N_1 q^4}{16 \pi \sqrt{2m^*} \kappa_0^2 \epsilon_0^2 }[ln(1+\gamma^2_{CW})]E^{-3/2}
    \label{equation:iis_weak}
\end{equation}

\begin{equation}
    \gamma_{CW} = \frac{4 \pi \kappa_0 \epsilon_0 E}{q^2 N_1^{1/3}}
    \label{equation:gamma_cw}
\end{equation}

The strong IIS screening limit is obtained by treating the external potential as a $\delta$-function, and is given by \cite{lundstrom}:

\begin{equation}
    \frac{1}{\tau_m} = \frac{\pi N_1}{\hbar} \biggl( \frac{q^2 L_{TF}^2}{\kappa_0 \epsilon_0}\biggl)^2 g(E)
    \label{equation:iis_strong}
\end{equation}

\begin{figure}[h]
    \raggedright
    \includegraphics[scale=0.15]{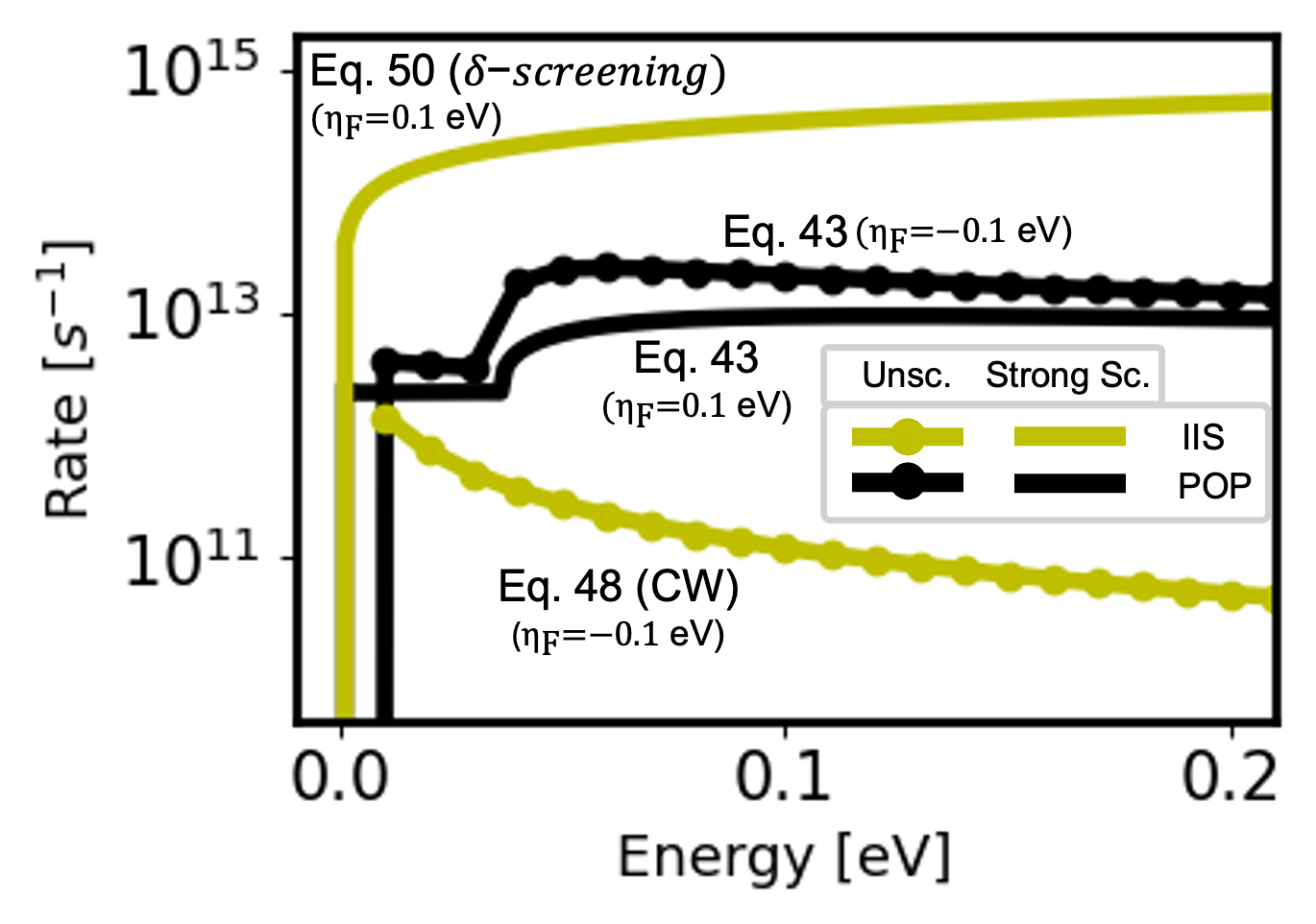}
    \caption{Comparison between weak and strong screening limits for IIS and POP scattering. We use $\eta_F = -0.1$ $eV$ for weak screening and $\eta_F = 0.1$ for strong screening.}
    \label{fig:fig4}
\end{figure}

For POP, taking the limit $L_{TF} \to \infty$, our newly derived screened expression (Eq. \ref{equation:pop_s_full}) would simplify into Eq. \ref{equation:pop_ns_full}, which is also the standard literature unscreened expressions \cite{lundstrom}. The obtained limits for POP and IIS are plotted in Fig. \ref{fig:fig4}. To approximate the limits of weak and strong screening, $\eta_F$ values of $-0.1$ and $0.1$ $eV$ were used, resulting in $L_{TF}$ values of 8.65 $nm$ and 0.85 $nm$ for POP (using $\kappa_\infty$ in the expression for $L_{TF}$) and 11.59 $nm$ and 1.14 $nm$ for IIS (using $\kappa_0$ in $L_{TF}$), respectively. Here, the differences in the variation between the POP and IIS rates are pronounced even more (black lines versus yellow lines). The POP rates are much closer together, whereas the IIS rates are orders of magnitude apart and diverge with energy. In the unscreened limit, the IIS scattering rates seem to decrease with energy (yellow line with symbols), much more compared to the POP rates (black line with symbols). In the strong screened limit, the IIS rate (solid yellow line) tends towards the energy dependence of the density of states ($\tau^{-1} \propto g(E) \propto E^{0.5}$), as in Eq. \ref{equation:iis_strong}. Interestingly, the strongly screened POP rate is lower compared to the unscreened rate, whereas the strongly screened IIS rate is much higher compared to the unscreened IIS rate. Since $L_{TF} \propto \sqrt{\frac{T\mathcal{F}_{1/2}(\eta)}{n\mathcal{F}_{-1/2}(\eta)}}$, as $n$ increases, the $L_{TF}^4$ that appears in both the POP and IIS expressions decreases fast, indicating that the process becomes more and more strongly screened. The $N_I$ proportionality of the IIS, however, increases its rates substantially. Also note that the Brooks-Herring model in our case does not enter the delta-function screening potential regime, and shows a decreasing trend with energy. At high $\eta_F$, the value of $L_{TF}$ is seen to stagnate (as shown in Figure \ref{fig:figS3}), preventing the model from entering the high-screened regime as seen in Figure \ref{fig:fig3}(b). Also note that the absolute values for POP and IIS presented in this example depend on the specific parameters used in each of the models. Thus, here we only consider the trends of the rates ($\kappa_{\infty}$ for example, only appears in POP). Our choice of $\kappa_{0}$/$\kappa_{\infty}$, however, resembles a strongly polar material.

\subsection{Electronic Transport Properties}

Once the scattering rates are obtained, the electronic transport properties, i.e. electrical conductivity $\sigma$ and mobility $\mu$, can be calculated. We use Boltzmann transport to evaluate these under low-field conditions \cite{Neophytou_2020}, for a single parabolic band material (with effective mass $m^* = 0.92$ $m_0$ in this work). The basic Kernel of the Boltzmann Transport Equation is the so-called Transport Distribution Function \cite{Neophytou_2020}: 

\begin{equation}
    \Xi(E) = \sum_{k}{\tau(k,E)v^2(k,E)g(k,E)}
    \label{equation:tdf}
\end{equation} 

The TDF incorporates the relaxation time ($\tau$), carrier velocity ($v$) and the density of states ($g$). The electrical conductivity, $\sigma$, and mobility, $\mu$, are given by: 

\begin{equation}
    \sigma = q^2 \int \Xi(E)(-\frac{df}{dE})dE
    \label{equation:sigma}
\end{equation} 

\begin{equation}
    \mu = \frac{\sigma}{q n}
    \label{equation:mu}
\end{equation} 

where $f$ is the equilibrium Fermi distribution.

\begin{figure}[h]
    \centering
    \includegraphics[scale=0.15]{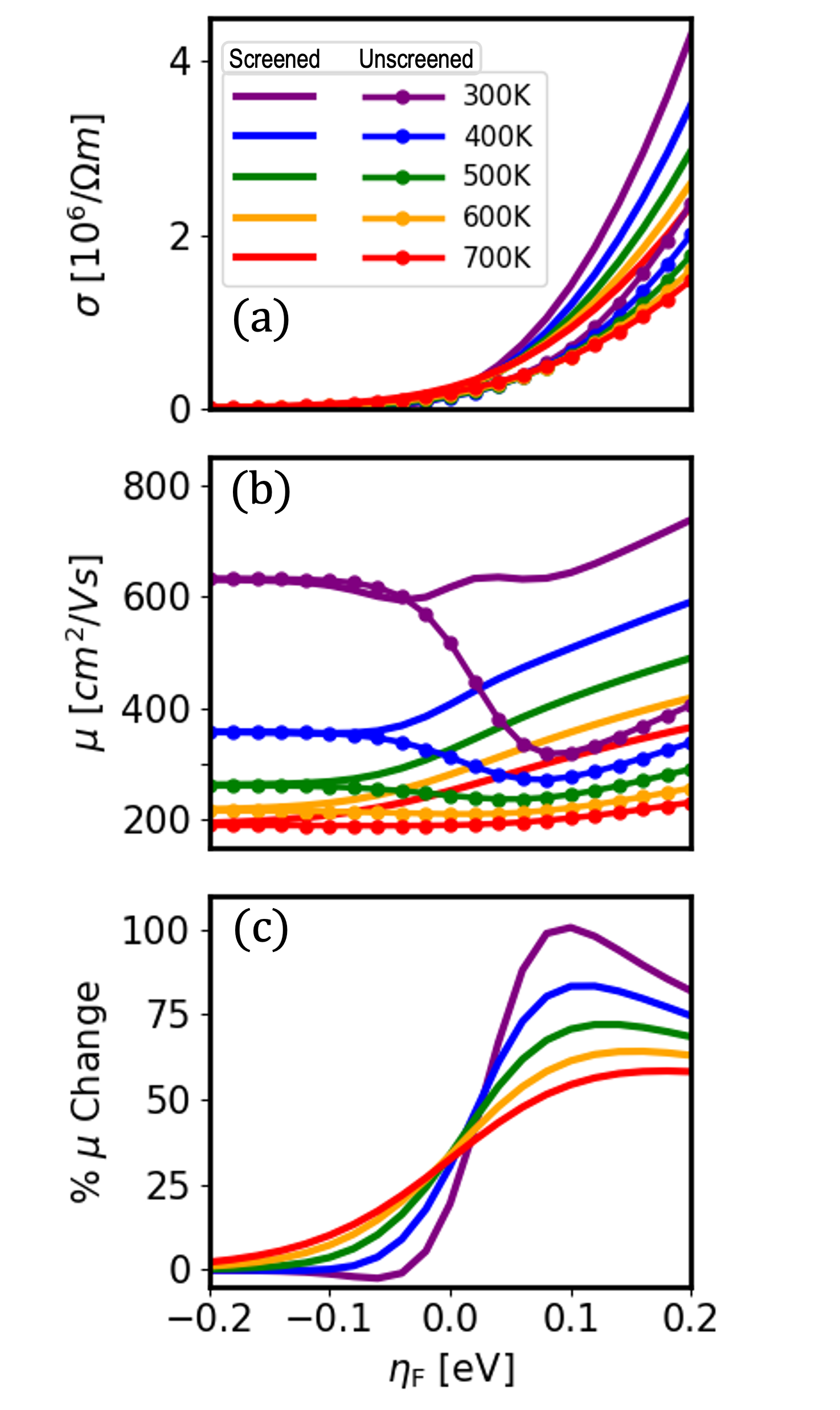}
    \caption{(a) Conductivity $\sigma$ versus Fermi level $\eta_F$ for multiple temperatures (T = $300$-$700^\circ K$) for screened (solid) and unscreened (solid-dot) POP scattering considerations. (b) Mobility $\mu$ versus $\eta_F$ for the data of panel (a). (c) The ratio of the screened to unscreened mobilities from panel (b) for all temperatures considered.}
    \label{fig:fig5}
\end{figure}

The results for electrical conductivity, $\sigma$, and charge mobility, $\mu$, using the derived scattering rates are plotted in Figure \ref{fig:fig5}, where panel \ref{fig:fig5}(a) plots the conductivity, and panels \ref{fig:fig5}(b) and \ref{fig:fig5}(c) plot the mobility and its percentage difference between the screened and unscreened scattering considerations. In panels \ref{fig:fig5}(a) and \ref{fig:fig5}(b) we show the screened cases by the solid lines, and the unscreened cases by the solid-dotted lines. We show results for multiple temperatures (all at and above room temperature). As expected, in the negative $\eta_F$ region (where the Fermi level is below the band and into the bandgap), in which the carrier density is low, the screened and unscreened treatments give identical results, with almost zero mobility variations (panel (c)). The mobility is clearly larger for lower temperatures, an expected behavior since the number of phonons is reduced with temperature. At higher $\eta_F$, where the carrier density increases and screening is strong, the screened scattering treatment produces larger values of $\sigma$ and $\mu$ for all temperatures (solid lines). The trend in Figure \ref{fig:fig5}(b) changes when the Fermi level overpasses the band edge, where for unscreened POP, $\mu$ has a flat/downward trend, while for screened POP it has an upward trend (note that here we only consider POP-limited results - if IIS is included, its downward trend will be emphasized). This leads to large discrepancies at all temperatures, as shown in Figure \ref{fig:fig5}(c), even up to 100$\%$. The mobility discrepancy between the screened and unscreened treatments increases with density ($\eta_F$) until $\eta_F = 0.1 eV$, while this discrepancy is stronger at lower temperatures. This is due to the screening length's, $L_{TF}$, dependence on density and temperature. As seen in Eq. \ref{equation:l_tf}, the screening length $L_{TF}$ is proportional to $\sqrt{T/n}$. Thus, at low T and large $n$, $L_{TF}$ is smaller, hence screening is stronger and the scattering rates are reduced (see next section for more elaborative discussion on the influence of temperature and density). This increases the transport properties $\sigma$ and $\mu$ compared to the unscreened considerations. Thus, using the screened POP scattering rate equation is indeed essential to capture the effect of POP correctly, particularly for high carrier densities and room or lower temperatures.

\subsection{Scattering Exponents}

\begin{figure}[h]
    \centering
    \includegraphics[scale=0.13]{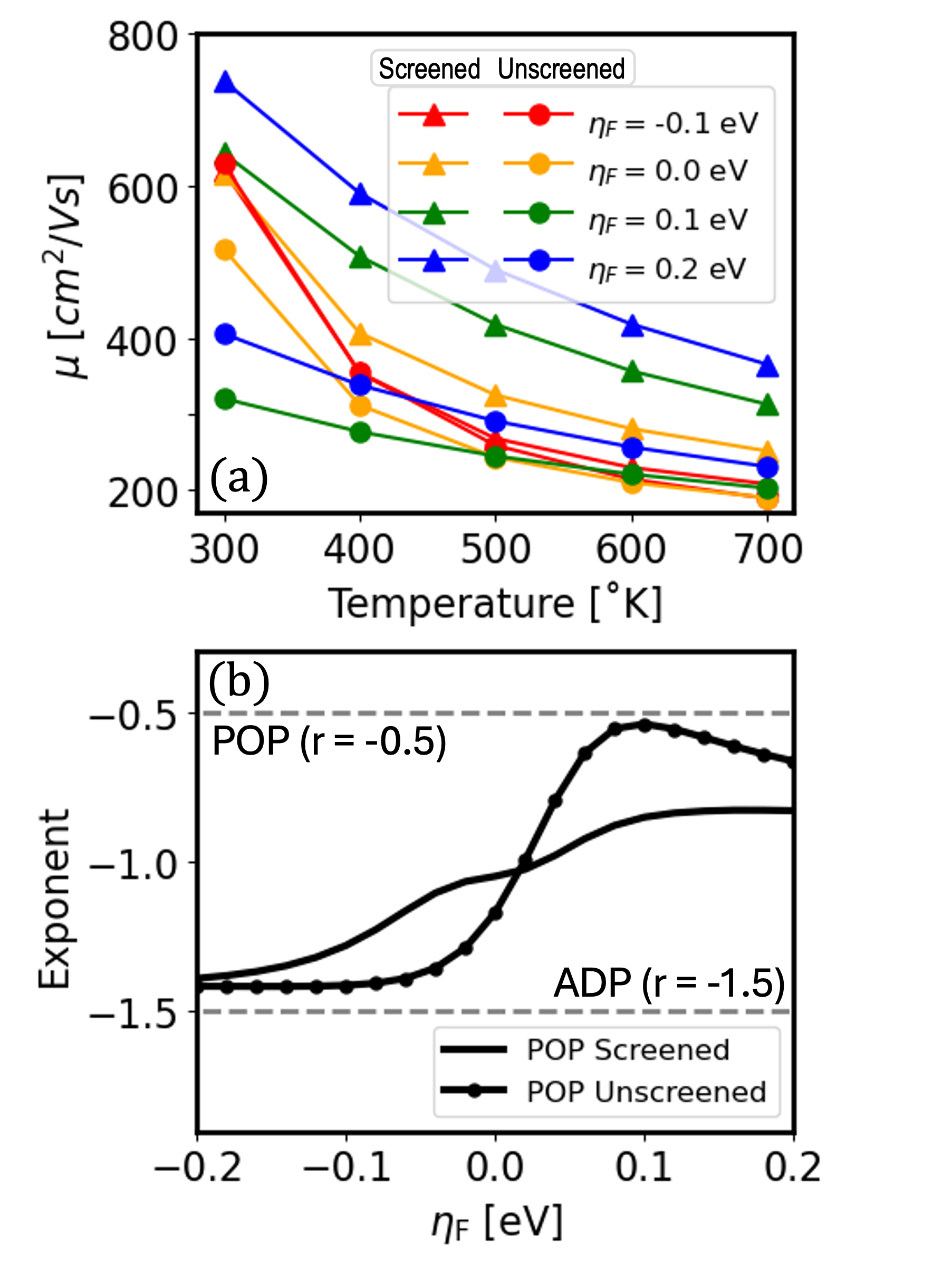}
    \caption{(a) Mobility $\mu$ against Temperature [K] for multiple values of $\eta_F$, for screened (solid-triangle) and unscreened (solid-dot) POP scattering. (b) The temperature dependent exponent ($\mu \propto T^r$) that describes the data in panel (a) versus Fermi Level $\eta_F$. Standard literature values for the exponents of POP and ADP are shown.}
    \label{fig:fig6}
\end{figure}

The exponential temperature dependence of mobility, in the form $\mu \propto T^r$, where $r$ is referred to as the scattering exponent, is an important way to provide insight into the dominant scattering mechanism in a material when analyzing experimental data. Using the obtained equations, Fig. \ref{fig:fig6}(a) plots the mobility, $\mu$, versus temperature for various values of $\eta_F$, for both the screened and unscreened cases. Figure \ref{fig:fig6}(b) plots the characteristic exponents extracted from Fig. \ref{fig:fig6}(a). The standard literature value for unscreened POP is $r = -0.5$ \cite{Ehrenreich,Ehrenreich_1957,ganose2,Petritz_Scanlon_1955}. This is what is observed at the high $\eta_F$ region for the unscreened model in Fig. \ref{fig:fig6}(b). At low $\eta_F$, however, the unscreened model maintains a low exponent of $r=-1.4$ up until $\eta_F = 0 eV$. The screened model starts at similar values, then increases linearly until it reaches $r = -0.8$ and saturates. The values given by POP under screening are between the literature value reported for acoustic deformation potential (ADP) scattering $(r = -1.5)$ and POP scattering. Indeed, both first principle studies and experimental works have shown that polar-dominated materials can have temperature dependencies close to the regular deformation potential scattering \cite{ganose, Cao_Querales-Flores_Murphy_Fahy_Savić_2018, Ma_Chen_Li_2018}, which validates our work as well. Thus, care needs to be taken in evaluating experimental data, since it will not be easy to distinguish between POP and non-polar deformation potential temperature trends from mobility measurements.

\section{Validity of the Quasi-Static Screening Method}

For the electron cloud to screen the electron - LO polar phonon interaction effectively, two conditions must be met. The first is the presence of an electron cloud, whose plasma frequency $\omega_{pl}$ must be greater than the LO phonon frequency $\omega_{ph}$, such that the plasma can respond fast enough and screen the electric field from the LO phonon interaction \cite{Ridley_2013, Ren2020}. The plasma frequency is given by:

\begin{equation}
    \omega_{pl}^2 = \frac{n q^2}{m^*\epsilon_0\kappa_\infty}
    \label{equation:omega_pl}
\end{equation}

where $n$ is the charge density, $m^*$ is the effective mass and $\kappa_\infty$ is the high frequency dielectric constant of the material. Since the plasma frequency increases with the square root of the density, screening is stronger at high densities. This is as expected, and what our results in Fig. \ref{fig:fig3} show (despite the fact that the plasma frequency never enters the derived formalism). Note that this is the reason why typically POP screening is not considered at low densities, as the low plasma frequency will not provide effective screening.

\begin{figure}[h]
    \centering
    \includegraphics[width=0.75\linewidth]{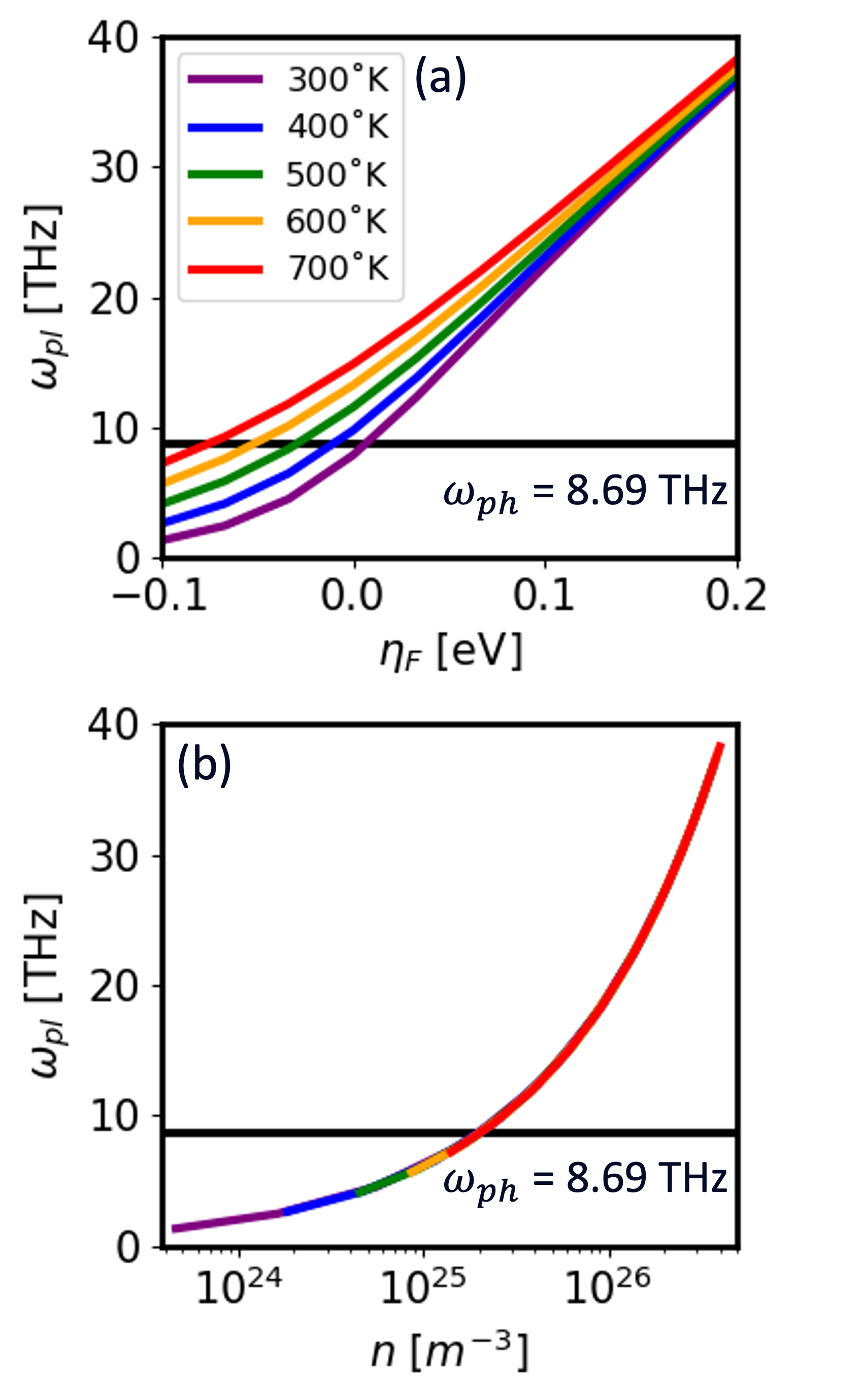}
    \caption{Colored lines: plasma frequency $\omega_{pl}$ against (a) $\eta_F$ and (b)  charge density $n$ for multiple temperatures. Black line: phonon frequency (8.69 THz) which is assumed to be constant throughout the manuscript.}
    \label{fig:figS1}
\end{figure}

Taking our work's example material NbFeSb, which is strongly polar, we compute the plasma frequency $\omega_{pl}$ and plot it against the reduced Fermi level $\eta_F$ and density $n$, for several temperatures, as shown in Figure \ref{fig:figS1}. The black horizontal line shows the LO phonon frequency used, $\omega_{ph}$, which corresponds to the material's LO frequency of 8.69 THz and energy of $36$ $meV$. Also note that the plasma frequency is typically given in units of [rad/s], thus we converted it to THz by dividing Eq. \ref{equation:omega_pl} by 2$\pi$. In Figure \ref{fig:figS1}(a), for large $\eta_F$ the plasma frequency is significantly higher compared to the LO phonon frequency for all temperatures, which can be identified as a regime of strong screening. In the region of lower $\eta_F$ (from -0.1 up to 0 eV) the plasma frequency becomes lower than the LO phonon frequency, especially for the lower temperatures, corresponding to partial or weak screening. Strictly speaking, for these cases, our screening formalism cannot apply, since the prerequisite of a fast enough electron plasma is not met. However, our formalism indicates that scattering is indeed weak in that regime, as shown in Figure \ref{fig:fig3}(a), where the derived screening equation result approximates the unscreened result very closely. Thus, both the plasma and the screened equation we derived, point towards the same conclusion, namely that screening is weak for low densities (in other words our screened formalism cannot be applied for low densities, but at low densities screening is weak anyway). Beyond $\eta_F > 0 eV$, it holds that $\omega_{pl} > \omega_{ph}$, and screening effects are strong and our formalism can be applied. However, the density and $\eta_F$ that determine the crossing of $\omega_{pl} > \omega_{ph}$, are material and temperature dependent. Thus, the density and $\eta_F$ region of validity also changes. For example, for the same carrier density and plasma frequency, lower effective mass materials and lower temperatures will require higher $\eta_F$, which reduces the $\eta_F$ range of validity in some cases. We show various material examples in Appendix \ref{chapter:validity_ex}, indicating this effect for low densities and temperatures.

In Fig. \ref{fig:figS1}(b) we show the same data, but now plotted as a function of density, rather than Fermi level. In this case the plasma frequency trends for different temperatures largely overlap. At density of around $2.00\times10^{25} m^{-3}$ the plasma frequency crosses the phonon frequency, clearly showing that there is a density boundary level that separates strong and weak screening regimes.

Any time-dependence of the scattering potential is assumed to be slow enough for the electron gas to respond and form a screening pattern. In other words, the screening has been assumed to be what it would be for a static potential. In the most general scenario, dynamic screening and plasma effects become inextricably mixed and must be treated together (electrons move under the influence of the electric field, which is the source of scattering, but the field also moves in time itself, thus the process becomes dynamic). In effect, we deal with coupled plasmon/polar-LO-phonon modes.\cite{Ridley_2013,Ren2020}.

Once the prerequisite of a fast oscillating plasma frequency and adequate density is available for screening, the second condition that needs to be fulfilled for effective screening of the LO polar phonon interaction by the electron cloud, is the relative timescales of the electron motion compared to the interaction motion. Note that the electrons do not need to follow the mechanical movement of the ions themselves, but the oscillation of the electric field that is produced by the phonon wave, which creates the long-range Coulomb field that electrons screen. This macroscopic electric field moves at the same speed as the phonon wave, meaning that it propagates at the LO phonon’s phase velocity ($\omega/q$), which describes the motion of the individual wave crests of the field. Thus, this determines how fast the phonon's electric field propagates through the material. Screening occurs when the electrons can respond quickly enough and rearrange themselves to shield this long range oscillating macroscopic electric field, and this is determined by their thermal velocity.

\begin{figure}[h]
    \centering
    \includegraphics[width=0.95\linewidth]{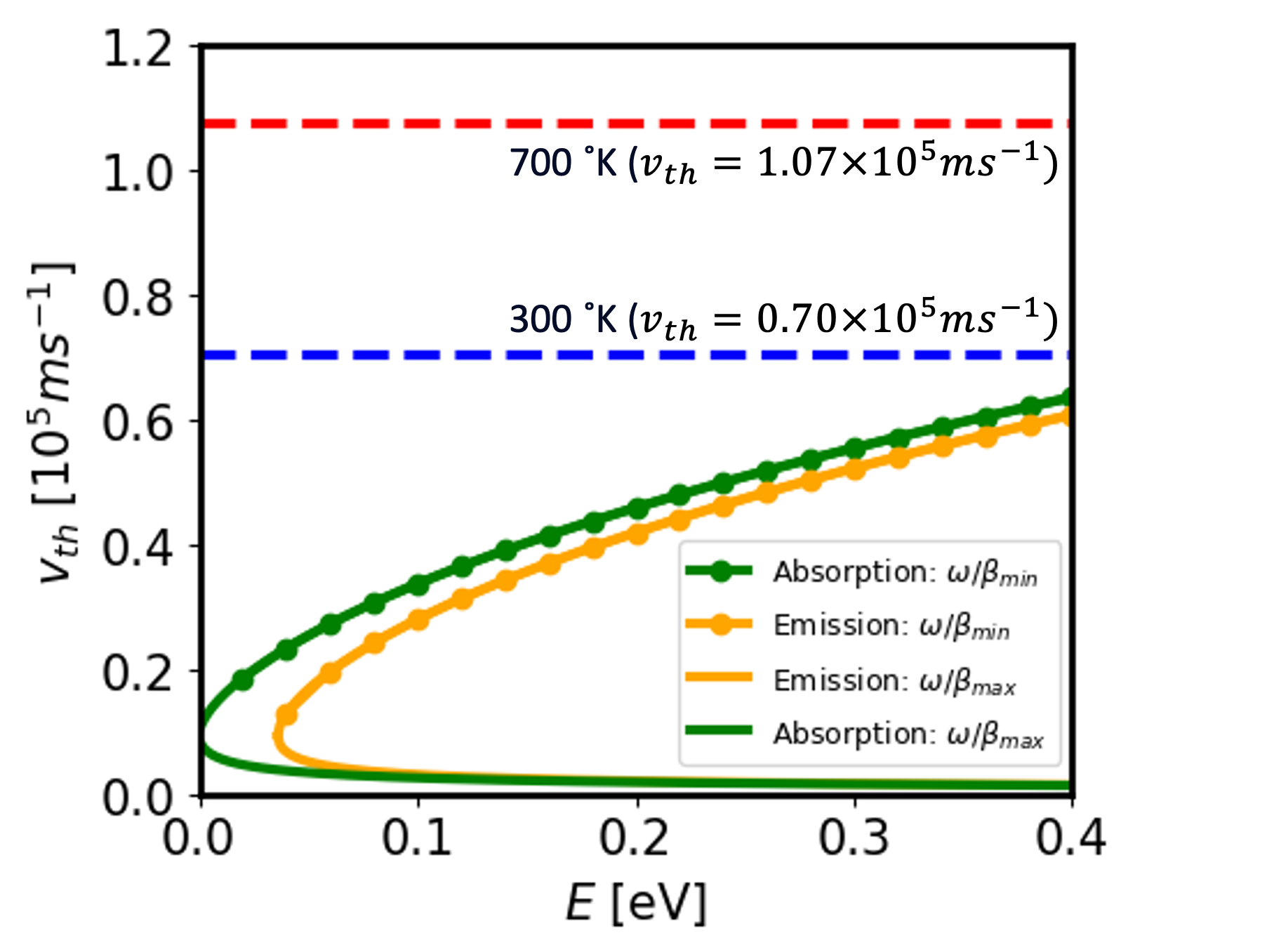}
    \caption{Solid lines show phase velocity $\omega/\beta$ against energy $E$ for emission (orange) and absorption (green) mechanism. Each color has 2 lines, which each show the corresponding velocities for $\beta_{max}$ (solid) and $\beta_{min}$ (dotted). Dashed lines show the thermal velocity at temperatures of $300^{\circ}K$ (blue) and $700^{\circ}K$ (red).}
    \label{fig:figS2}
\end{figure}

The electron thermal velocity $v_{th}$ is given by \cite{Baumjohann2021} :

\begin{equation}
    v_{th} = \sqrt{\frac{k_BT}{m^*}}
\end{equation}

which is obtained by equating the average thermal energy of the carriers, $\frac{1}{2}k_BT$, to their kinetic energy, $\frac{1}{2}m^*v_{th}^2$, where $m^*$ is the effective mass of the charge carrier. Typically, $v_{th}$ is larger than the phonon phase velocity. To show the corresponding comparison, which will also provide the region of validity of our formalism, in Figure \ref{fig:figS2} we plot the extend of the phase velocities ($\omega_{ph}/\beta$) of all phonons that participate in the POP scattering process, and compare that to $v_{th}$ at $300^\circ K$ and $700^\circ K$. For this, we plot the phase velocities corresponding to $\beta_{min}$ and $\beta_{max}$ for both emission and absorption (defined as $\omega_{ph}/\beta$). This essentially defines the outer boundaries of the phase velocities of the phonons that take part in the scattering processes. Note that the frequency $\omega_{ph}$ here is expressed in THz (8.69 THz in this case), in order for $\omega_{ph}/\beta$ to be compared to a velocity. Clearly, the thermal velocity of electrons even down to $300^\circ K$ is larger compared to the phase velocity of the participating phonons, thus, at those temperatures our results will be fully valid. At lower temperatures, however, the thermal velocity will be lower, and computing screening at the higher energy regions might not be applicable using static screening methods. Note, however, that typically the Fermi level is placed around the band edge for most applications, and at low temperatures only low energy electrons near the band edge participate in transport, which have lower phase velocities, thus somewhat enhancing the applicability of the formalism even at lower temperatures.

At very high temperatures, on the other hand, the thermal velocity of the electrons can be so high that electrons move very fast. The presence of these fast electrons (hot carriers), actually weaken screening, because electrons move too chaotically to effectively neutralize the phonon-induced electric field. So, electrons need to move fast enough compared to the phonon phase velocity, but not so fast that they cannot localize and screen the oscillations. 

Thus, for large screening strength you need large plasma frequency, which determines the screening strength, and moderate thermal velocity, for electrons to respond fast enough, but not diffuse too much. If the plasma frequency is low, even if thermal velocity is large, screening is not effective since there are no electrons to screen the electric field. Thus, the plasma frequency is the predominant factor, while thermal velocity dictates how the existing electrons will dynamically react to the field oscillations. Note that the need for large density to have fast electron plasma oscillations and the need for moderate/smaller temperatures to allow the electrons to localize around the field oscillation, is captured in the screening length, which is proportional to $L_{TF}\propto\sqrt{T/n}$, and the smaller it is, the stronger the screening. 

\section{Conclusion}

In conclusion, in this work we presented a new analytical mathematical formula to calculate POP scattering rates including the effect of screening for a parabolic band. We have shown that for strong screening, the new expression produces largely different values for the electrical conductivity, $\sigma$, and mobility, $\mu$, compared to the usual unscreened treatment, leading up to a deviation of 100$\%$. This difference is particularly enhanced for high densities and lower temperatures. The temperature-dependent exponent of the mobility under the POP scattering limiting process is also found to be different for the two treatments. The screened POP model produces exponents that vary between $r=-1.5$ and $r=-0.8$, in some regions very similar to what acoustic deformation potential scattering is believed to result to. Finally, we present an analysis for the conditions under which the validity of this quasi-static screening approach holds in relation to the electron plasma frequency and their thermal velocity, which are different for different materials and temperatures.

\begin{acknowledgments}
This work was supported by EPSRC Centre for Doctoral Training in Modelling of Heterogeneous Systems (HetSys, Project Reference No. 2887654) and from the UK Research and Innovation fund (project reference EP/X02346X/1).
\end{acknowledgments}

\section*{Data Availability}
 The data are available from the authors upon reasonable request.

\setcounter{section}{0}
\setcounter{equation}{0}
\setcounter{figure}{0}

\renewcommand{\thesection}{}
\renewcommand{\theequation}{\Alph{subsection}\arabic{equation}}
\renewcommand{\thetable}{\Alph{subsection}.\arabic{table}}
\renewcommand{\thefigure}{\Alph{subsection}.\arabic{figure}}
\renewcommand{\thesubsection}{\Alph{subsection}}

\section*{Appendices}

\subsection{Effective charges}
\label{chapter:eff_charge}

\begin{table}[h!]
\centering
{%
\begin{tabular}{|l|l|l|l|}
\hline
\multicolumn{1}{|c|}{\textbf{Material}} & \multicolumn{1}{c|}{\textbf{$q^*$}} \\ \hline
AlSb                                    & 0.48                             \\ \hline
GaN                                     & 1.44                             \\ \hline
GaP                                     & 0.57                             \\ \hline
GaAs                                    & 0.33                             \\ \hline
InP                                     & 0.69                             \\ \hline
InAs                                    & 0.57                             \\ \hline
InSb                                    & 0.28                             \\ \hline
ZnO                                     & 1.1                              \\ \hline
ZnS                                     & 0.84                             \\ \hline
ZnSe                                    & 0.72                             \\ \hline
NbFeSb                                  & 0.59                             \\ \hline
\end{tabular}%
}
\caption{The effective charge $q^*$ for common materials. Values obtained from \cite{Callen_1949,Lawaetz}. See references for relevant list of material parameters leading to these values.}
\label{table:ef_charge}
\end{table}

For NbFeSb, we used values of $a_0 = 0.42$ nm, $M = 4.5\times10^{-26}$ kg, $\kappa_0 = 42.83$, $\kappa_\infty = 23.91$, $\hbar\omega_0 = 36$ $meV$ \cite{Jain2013}. The volume of the crystal was calculated as $2a_0^3$, as done in \cite{Callen_1949,Lawaetz}.

\setcounter{equation}{0}
\setcounter{figure}{0}
\subsection{Derivation of generalized screening term}
\label{chapter:screening_term}

To reach the screening term, $\frac{\beta^2}{\beta^2+\frac{1}{L_{TF}^2}}$, which modifies the perturbing potential, the original derivation goes back to the work of Ehrenreich \cite{Ehrenreich1959}. Here we partially follow a newer work by Ridley, Ch. 9.5 \cite{Ridley_2013}, who uses the theory of the Lindhard dielectric function. When considering Coulombic screening, the total potential $\phi_T$ that electrons experience, which includes screening, can be expressed assuming linear response as a function of the unscreened potential $\phi_u$ and the dielectric permittivity function $\epsilon(\beta,\omega)$, as:

\begin{equation}
    \phi_T = \frac{ \phi_u}{\epsilon(\beta,\omega)}
    \label{equation:phi_s_2}
\end{equation}

We need to describe the response of the electronic system to a weak wave-like potential. Starting from the unscreened potential, we first assume that it has the form:

\begin{equation}
    \phi_u(\beta,\omega) = V_ue^{i(\beta\cdot r-\omega t)}
    \label{equation:phi_u}
\end{equation}

where $V_u$ is its amplitude and $\beta$ is its wavenumber. 

We then assume that the response is linear such that the perturbed electron density follows the variation in the potential energy, $-q\phi_u$, as: 

\begin{equation}
    n_s = F(\omega,\beta)(-q\phi_u)
    \label{equation:n_s}
\end{equation}

where $-q$ is the electronic charge and $F(\beta,\omega)$ is the characteristic density response function. Note that the units of $F(\beta,\omega)$ are $[s^2kg^{-1}m^-5]$. The variation of density gives rise to a screening potential $\phi_s$, which also follows the unscreened potential and the density variation as:

\begin{equation}
    \phi_s(\beta,\omega) = V_{s}e^{i(\beta\cdot r-\omega t)}
    \label{equation:phi_s}
\end{equation}

where $V_s$ is the amplitude of the oscillation. Using Poisson's equation for the screening potential ($-\epsilon_0\kappa\nabla^2\phi_s = -qn_s$), we obtain:

\begin{equation}
    \phi_s = \frac{-qn_s}{\epsilon_0 \kappa\beta^2}
\end{equation}

The screening potential $\phi_s$ can now be expressed in terms of $\phi_u$ by:

\begin{equation}
    \phi_s = \frac{-q}{\epsilon_0\kappa\beta^2}n_s = \frac{q^2}{\epsilon_0\kappa\beta^2}F(\omega,\beta)\phi_u
\end{equation}

The total potential, $\phi_T$, experienced by the electrons, is now the summation of the screened and unscreened terms as:

\begin{equation}
    \phi_T = \phi_u + \phi_s
    \label{equation:phi_T_sum}
\end{equation}

Substituting for $\phi_s$, an expression for $\phi_T$ is obtained as:

\begin{equation}
    \phi_T = \bigg(1 + \frac{q^2}{\epsilon_0\kappa \beta^2}F(\omega,\beta)\bigg)\phi_u
    \label{equation:phi_T}
\end{equation}

We can then define the linear response of the density to this potential as:

\begin{equation}
    n_s = G(\beta,\omega)(-q \phi_T)
    \label{equation:phiT_response}
\end{equation}

where $G(\beta,\omega)$ is the response function. This is calculated using the Liouville equation for the time evolution of the quantum mechanical density matrix\cite{Ridley_2013}, and is given by:

 \begin{equation}
    G(\beta,\omega) = \lim_{\alpha\to0}\sum_k \frac{f_{k+\beta} - f_{k}}{E_{k+\beta}-E_{k}-\hbar\omega+i\hbar\alpha}
    \label{equation:g_response}
\end{equation}

where $f_k$ is the distribution function for state $k$, $E_k$ is the energy of state $k$, $\omega$ is the phonon frequency and $\alpha$ is an adiabatic tuning term. Using the linear relationship between $\phi_T$ and $\phi_u$ (Eq. \ref{equation:phi_T}), we can define $n_s$ (from Eq. \ref{equation:n_s}) in terms of $\phi_T$ as:

\begin{equation}
    n_s = \frac{F(\beta,\omega)}{1 + \frac{q^2}{\epsilon_0\kappa\beta^2}F(\beta,\omega)}(-q\phi_T)
    \label{equation:phiT_F_response}
\end{equation}

By equating the right hand sides of Eq. \ref{equation:phiT_response} and Eq. \ref{equation:phiT_F_response}, we can express $F(\beta,\omega)$ in terms of $G(\beta,\omega)$. With this, Eq. \ref{equation:phi_T} can be rewritten in terms of $G(\beta,\omega)$. After simplifications, we can express $\phi_T$ as a function of $G(\beta,\omega)$, reaching:

\begin{equation}
    \phi_T = \frac{\phi_u}{1 - \frac{q^2}{\epsilon_0\kappa\beta^2}G(\beta,\omega)} = \frac{\phi_u}{\epsilon(\beta,\omega)}
\end{equation}

which gives the Lindhard formula \cite{Frhlich1954}:

\begin{equation}
    \epsilon(\beta,\omega) = 1 - \frac{q^2}{\epsilon_0\kappa\beta^2}\lim_{\alpha\to0}\sum_k \frac{f_{k+\beta} - f_{k}}{E_{k+\beta}-E_{k}-\hbar\omega+i\hbar\alpha}
\end{equation}

Taking ($\alpha, \omega) \to 0$ (static approximation), the equation is simplified to:

\begin{equation}
    \epsilon(\beta) = 1 - \frac{q^2}{\epsilon_0\kappa\beta^2}\sum_k \frac{f_{k+\beta} - f_{k}}{E_{k+\beta}-E_{k}}
    \label{equation:lind_simplified}
\end{equation}

The numerator can be approximated at first order to:

\begin{equation}
    f_{k+\beta} - f_k = f_k + \beta\frac{\partial f_k}{\partial k} + ... - f_k \approx \beta\frac{\partial f_k}{\partial k}
\end{equation}

while the denominator can be approximated to:

\begin{equation}
    E_{k+\beta} - E_k = \frac{\hbar^2}{2m}(k^2 + 2k\cdot \beta + \beta^2) - \frac{\hbar k^2}{2m} \approx \frac{\hbar^2 k\cdot\beta}{m}
\end{equation}

Substituting the approximations above into the Lindhard function (Eq. \ref{equation:lind_simplified}) gives:

\begin{equation}
    \epsilon(\beta) = 1 - \frac{q^2}{\epsilon_0\kappa\beta^2}\sum_k \frac{\beta\frac{\partial f_k}{\partial k}}{\frac{\hbar^2 k\cdot\beta}{m}}
    \label{equation:lind_simplified_2}
\end{equation}

Assuming thermal equilibrium is achieved, we can simplify the numerator as follows. We start with the Fermi-Dirac distribution:

\begin{equation}
    f_k = \frac{1}{1 + e^{(E_k - \mu)/k_BT}}  = \frac{1}{1 + e^{(\hbar^2k^2/2m-\mu)/k_BT}}
\end{equation}

where $E_K$ is the energy dispersion, $\mu$ is the chemical potential (interchangeably referred to as Fermi level elsewhere in this work), $T$ is the temperature, and $m$ is the mass of the particle. Then the fraction $\frac{\partial f_k}{\partial k}$ can be expressed as:

\begin{equation}
    \frac{\partial f_k}{\partial k} = \frac{-1/(k_BT)}{(1 + e^{(E_k - \mu)/k_BT})^2} (\hbar^2k/m)
\end{equation}

One can easily see that the first fraction of the equation above is equal to (-$\frac{\partial f_k}{\partial \mu}$).

Hence the total fraction is given by:

\begin{equation}
    \sum_k\beta\frac{\partial f_k}{\partial k} = -\sum_k\beta\frac{\partial f_k}{\partial \mu}\frac{\partial E_K}{\partial k} = -\sum_k\beta\frac{\hbar^2k}{m}\frac{\partial f_k}{\partial \mu}
\end{equation}

Substituting it back into Eq. \ref{equation:lind_simplified_2}, we can further simplify to obtain:

\begin{equation}
    \epsilon(\beta) = 1 + \frac{q^2}{\epsilon_0\kappa\beta^2}\sum_k \frac{\partial f_k}{\partial \mu}
    \label{equation:lind_simplified_3}
\end{equation}

Taking the partial derivative before the summation and using $\sum f_k = n$, we obtain the final expression for $\epsilon(\beta)$ as:

\begin{equation}
    \epsilon(\beta) =1 + \frac{q^2}{\epsilon_0\kappa\beta^2} \frac{\partial n} {\partial\mu}  = 1 + \frac{1}{\beta^2 L^2}
    \label{equation:epsilon_lindhard}
\end{equation}

where $L$ is the generalized screening length ($L_{TF} = \sqrt{\frac{\epsilon_0\kappa}{q^2}(\frac{\partial n}{\partial \mu})^{-1}}$), which the inverse of the Thomas-Fermi wavevector. Finally, substituting Eq. \ref{equation:epsilon_lindhard} into Eq. \ref{equation:phi_s_2} gives the characteristic screening term as:

\begin{equation}
    \phi_s = \phi_u \Bigg(\frac{1}{1 + \frac{1}{\beta^2 L^2}}\Bigg) = \phi_u \Bigg(\frac{\beta^2}{\beta^2 + \frac{1}{L^2}}\Bigg)
    \label{equation:phi_s_final}
\end{equation}

\setcounter{equation}{0}
\setcounter{figure}{0}
\subsection{Screening length}
\label{chapter:l_tf}

\indent The origin of both $L_{TF}$ and $L_D$ comes from Thomas-Fermi theory, which gives the expression for the Thomas-Fermi wavevector (inverse screening length) as \cite{Ashcroft1976}:

\begin{equation}
    \kappa = \sqrt{\frac{q^2}{\kappa \epsilon_0}\frac{\partial n}{\partial\mu}}
    \label{equation:general_k}
\end{equation}

where $q$ is the electronic charge, $\epsilon_0$ is the dielectric constant, $\kappa$ is the relative dielectric constant, $n$ is the charge density and $\mu$ is the chemical potential (Fermi energy). The Thomas-Fermi length, which is the inverse of the wavevector, is derived in Appendix \ref{chapter:screening_term} using the Lindhard dielectric function \cite{Ridley_2013}.

\indent The theories of Thomas-Fermi screening length and Debye screening length diverge here, on how the partial derivative $\frac{\partial n}{\partial \mu}$ is evaluated. 

\indent The Debye treatment follows Boltzmann statistics for the electron distribution, such as the density is proportional to:

\begin{equation}
    n \propto e^{\frac{\eta_F}{k_BT}}
    \label{equation:n_tf}
\end{equation}

\indent where $\eta_F=E_F-E_C$ is the reduced fermi level. Taking the differential $\frac{\partial n}{\partial \mu}$ gives:

\begin{equation}
    \frac{\partial n}{\partial \mu} = \frac{n}{k_BT}
    \label{equation:partial_deb}
\end{equation}

Substituting Eq. \ref{equation:partial_deb} into Eq. \ref{equation:general_k}, we obtain the final expression for $L_{D}^2$ as:

\begin{equation}
    L_{D}^2 =  \frac{\kappa \epsilon_0 k_B T}{q^2n}
\end{equation}

The Thomas-Fermi screening length is obtained by using Fermi-Dirac statistics for the electron distribution, such that the electron density $n$ is given by:

\begin{equation}
    n = N_c \mathcal{F}_{1/2}(\eta)
    \label{equation:tf_charge}
\end{equation}

where $N_c$ is the effective density of states of the band, $\mathcal{F}_{1/2}(\eta)$ is the Fermi-Dirac integral of order $1/2$, and $\eta = \frac{\eta_F}{k_BT}$.
Evaluating the partial derivative by using the chain rule, we obtain the following expression:

\begin{equation}
    \frac{\partial n}{\partial{\mu}} = \frac{\partial \mathcal{F}_{1/2}(\eta)}{\partial \eta}\frac{\partial \eta}{\partial\mu} = N_C \mathcal{F}_{-1/2}(\eta)\frac{1}{k_BT}
\end{equation}

Substituting $N_C$ from Eq. \ref{equation:tf_charge} and rearranging, we obtain the final expression for $L_{TF}^2$ as:

\begin{equation}
    L_{TF}^2 =  \frac{\kappa \epsilon_0 k_B T}{q^2n}\frac{\mathcal{F}_{1/2}(\eta)}{\mathcal{F}_{-1/2}(\eta)}
    \label{equation:ld_full}
\end{equation}

\begin{figure}[h]
    \centering
    \includegraphics[width=0.7\linewidth]{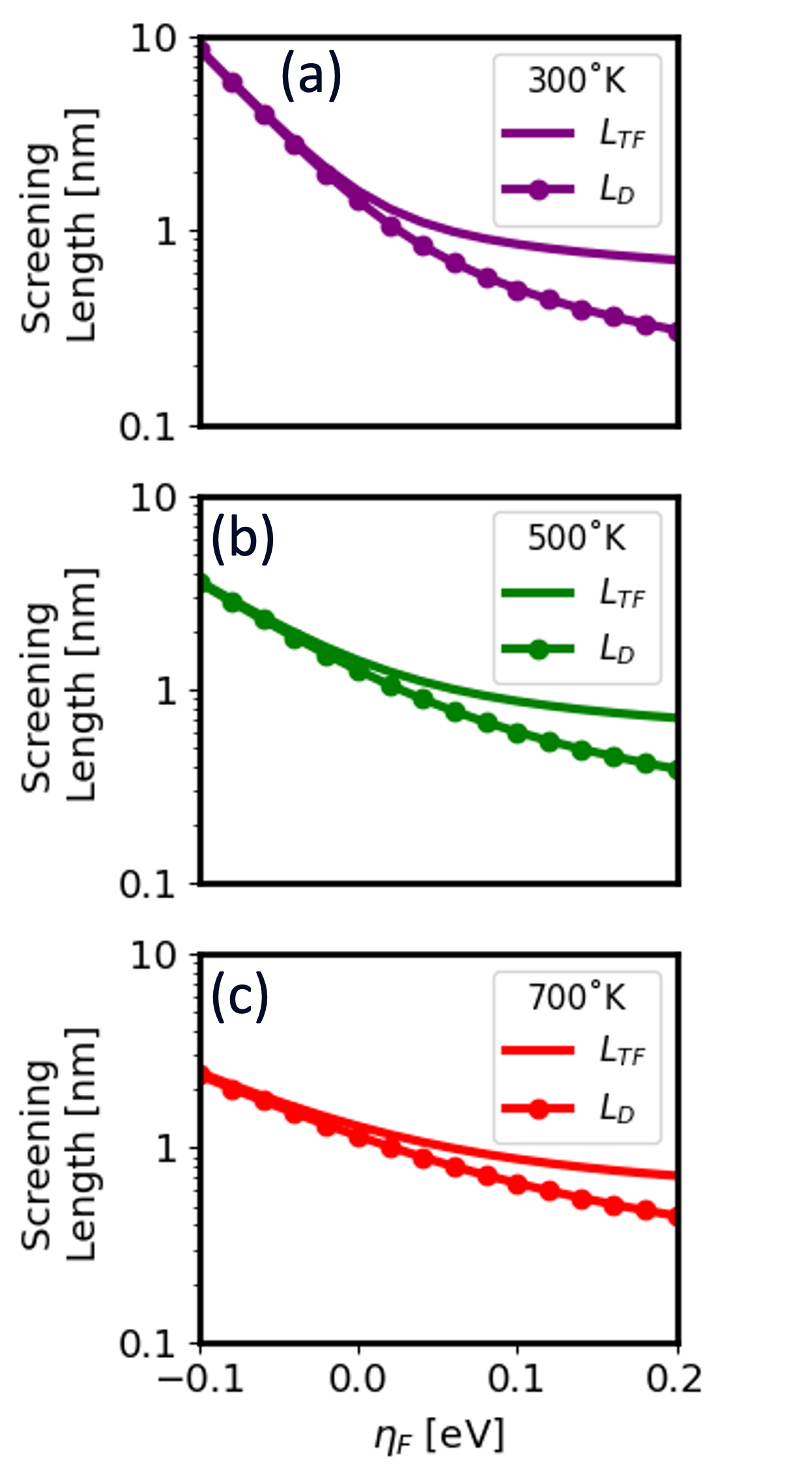}
    \caption{Screening lengths $L_{TF}$ (solid) and $L_{D}$ (solid-dotted) against $\eta_F$ for different temperatures: (a) $300^\circ K$, (b) $500^\circ K$, and (c) $700^\circ K$.}
    \label{fig:figS3}
\end{figure}

Figure \ref{fig:figS3} plots the screening lengths $L_D$ and $L_{TF}$ against $\eta_F$ for three different temperatures. In the non-degenerate regime $L_D$ and $L_{TF}$ converge, while in the degenerate regime, $L_{TF}$ deviates from $L_{D}$ due to the use of the Fermi-Dirac integrals in Eq. \ref{equation:ld_full}. $L_{TF}$ is larger as the density increases compared to $L_{D}$, indicating that the latter will overestimate screening and consequently the conductivity and mobility of the material.

\setcounter{equation}{0}
\setcounter{figure}{0}
\subsection{Integrals}
\label{chapter:integrals}
Solving for the integral in Eq. \ref{equation:int1}

\begin{equation}
    I_1 = \int \frac{\beta^5}{(\beta^2+\frac{1}{L_{TF}^2})^2}d\beta
\end{equation}

Solve the integration by using integration by parts

\begin{table}[h]
\centering
{%
\begin{tabular}{ll}
 $f = \beta^4$&   $g = \frac{-1}{2}\frac{1}{\beta^2 + \frac{1}{L_{TF}^2}}$ \\
 $f' = 4\beta^3$&  $g' = \frac{\beta}{(\beta^2+\frac{1}{L_{TF}^2})^2}$
\end{tabular}%
}
\end{table}

Substitute values of $f$, $g$, $f'$ and $g'$ into equation

\begin{equation}
    I_1 = \int fg'd\beta = fg - \int f'gd\beta
\end{equation}

\begin{equation}
    I_1 =  \frac{-\beta^4}{2(\beta^2+\frac{1}{L_{TF}^2})}-\int\frac{-2\beta^3}{\beta^2 +\frac{1}{L_{TF}^2}}d\beta
\end{equation}

Substitute $\beta = \frac{1}{L_{TF}}tan(\theta)$, where $\frac{d\beta}{d\theta} = \frac{1}{L_{TF}}sec^2(\theta)$.

\begin{equation}
    I_1 =  \frac{-\beta^4}{2(\beta^2+\frac{1}{L_{TF}^2})}-\int\frac{-2(\frac{tan(\theta)}{L_{TF}})^3}{\frac{tan^2(\theta)}{L_{TF}^2} +\frac{1}{L_{TF}^2}}\frac{sec^2(\theta)}{L_{TF}}d\theta
\end{equation}

\begin{equation}
    I_1 =  \frac{-\beta^4}{2(\beta^2+\frac{1}{L_{TF}^2})}-\frac{1}{L_{TF}^2}\int\frac{-2tan^3(\theta)sec^2(\theta)}{tan^2(\theta)+1}d\theta
\end{equation}

\begin{equation}
    I_1 =  \frac{-\beta^4}{2(\beta^2+\frac{1}{L_{TF}^2})}+\frac{2}{L_{TF}^2}\int tan^3(\theta)d\theta
\end{equation}

The integral of $tan^3(\theta)$ can be obtained from standard integral tables as:

\begin{equation}
    I_1 =  \frac{-\beta^4}{2(\beta^2+\frac{1}{L_{TF}^2})}+\frac{2}{L_{TF}^2}[\frac{sec^2(\theta)}{2}-ln|sec(\theta)|]
\end{equation}

Substitute $sec(\theta) = \sqrt{1+tan^2(\theta)} = \sqrt{1 + L_{TF}^2\beta^2}$ back to retrieve original variables to obtain:

\begin{equation}
    I_1 =  \frac{-\beta^4}{2(\beta^2+\frac{1}{L_{TF}^2})}+\frac{1}{L_{TF}^2}[1 + L_{TF}^2\beta^2 - ln(1+L_{TF}^2\beta^2)]
\end{equation}

as given in Eq. \ref{equation:int1_sol} of the main text.

\setcounter{equation}{0}
\setcounter{figure}{0}
\subsection{Regions of validity: examples}
\label{chapter:validity_ex}

\begin{figure}[h!]
    \centering
    \includegraphics[width=1\linewidth]{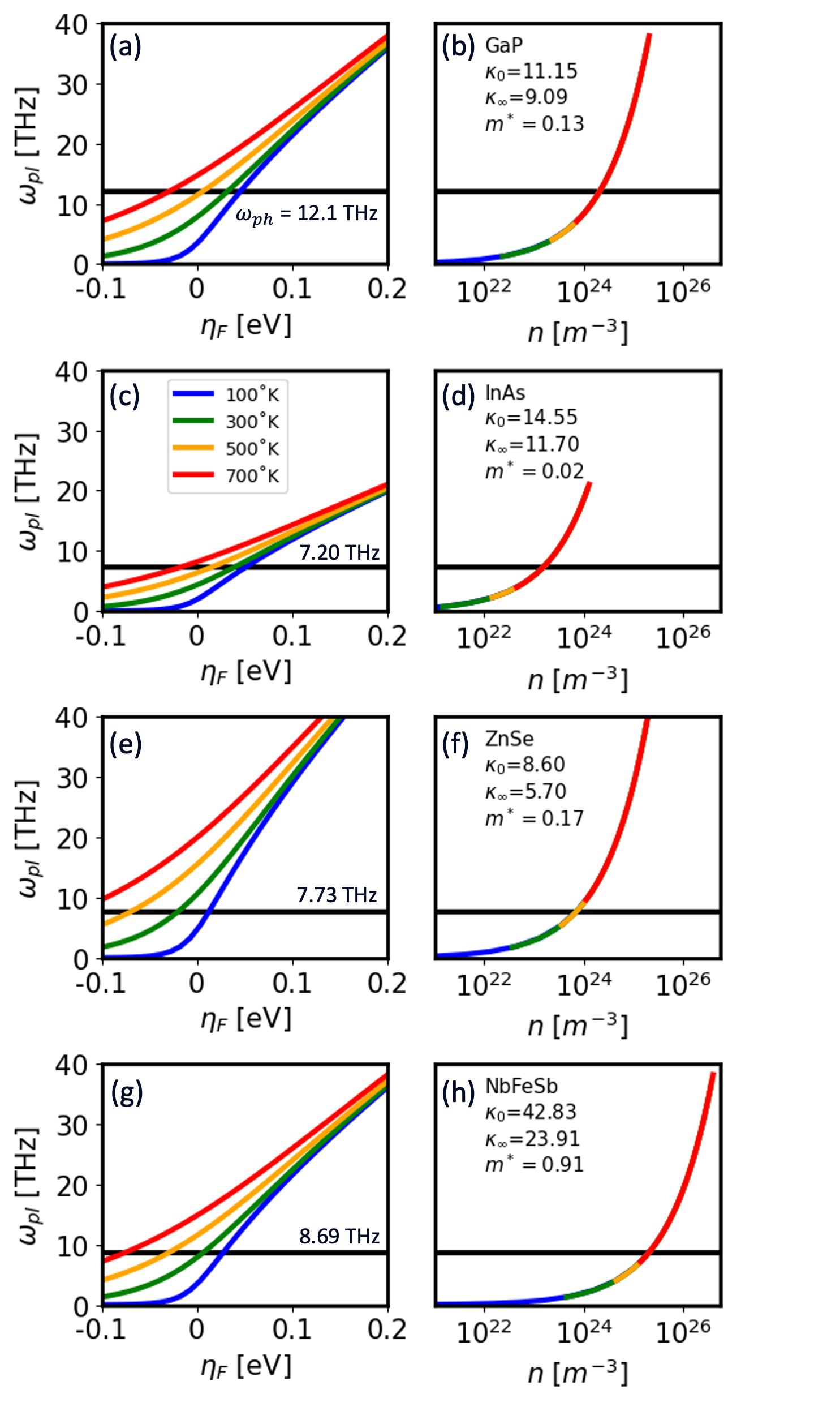}
    \caption{Colored lines: plasma frequency $\omega_{pl}$ against $\eta_F$ (left) and charge density $n$ (right) for multiple temperatures. Black flat line: LO phonon frequency for each material. The plasma frequency is plotted for four n-type materials: GaP (a-b), InAs (c-d) and ZnSe (e-f) and NbFeSb (g-h). Values obtained from Ref. \cite{Gbel1999,Lockwood2005,Marple1964,Wang1981,Adachi1991}}
    \label{fig:figS4}
\end{figure}

In Fig. \ref{fig:figS4} we considered four different materials (GaP, InAs, ZnSe, NbFeSb) ranked in order of ascending polarity ($\kappa_0/\kappa_\infty$). We plot the plasma frequency $\omega_{pl}$ against reduced Fermi level $\eta_F$ and charge density $n$ in the left and right columns, respectively.   

In the majority of cases, $\omega_{pl} > \omega_{ph}$ holds true approximately beyond $\eta_F > 0$ eV, suggesting that the quasi-static screening formalism can be applied. The carrier densities of validity also change for each material. Note that the crossing density point is the same for all temperatures for a given material, as the plasma frequency does not depend on temperature. For the lower temperature ($100^\circ K$) and the low mass material (InAs), the $\eta_F$ crossing point shifts slightly to the right. In particular, for InAs the crossing point density is the lowest compared to other materials due to its low effective mass.  

In general, materials with light effective mass $m^*$ and high $\kappa_\infty$ (e.g. InAs) have more restricted $\eta_F$ validity regions. However, due to the lighter effective mass $m^*$, the crossing point charge density $n$ decreases, and hence the screening length $L_{TF}$ increases as well. This reduces the degree of screening experienced at a given $\eta_F$, thus screening is weaker at those higher $\eta_F$ anyway. This suggests that in such cases, despite the fact that the validity of static screening is somewhat reduced compared to high effective mass materials, it is less relevant because screening is weaker.  

\begin{figure}[h!]
    \centering
    \includegraphics[width=0.9\linewidth]{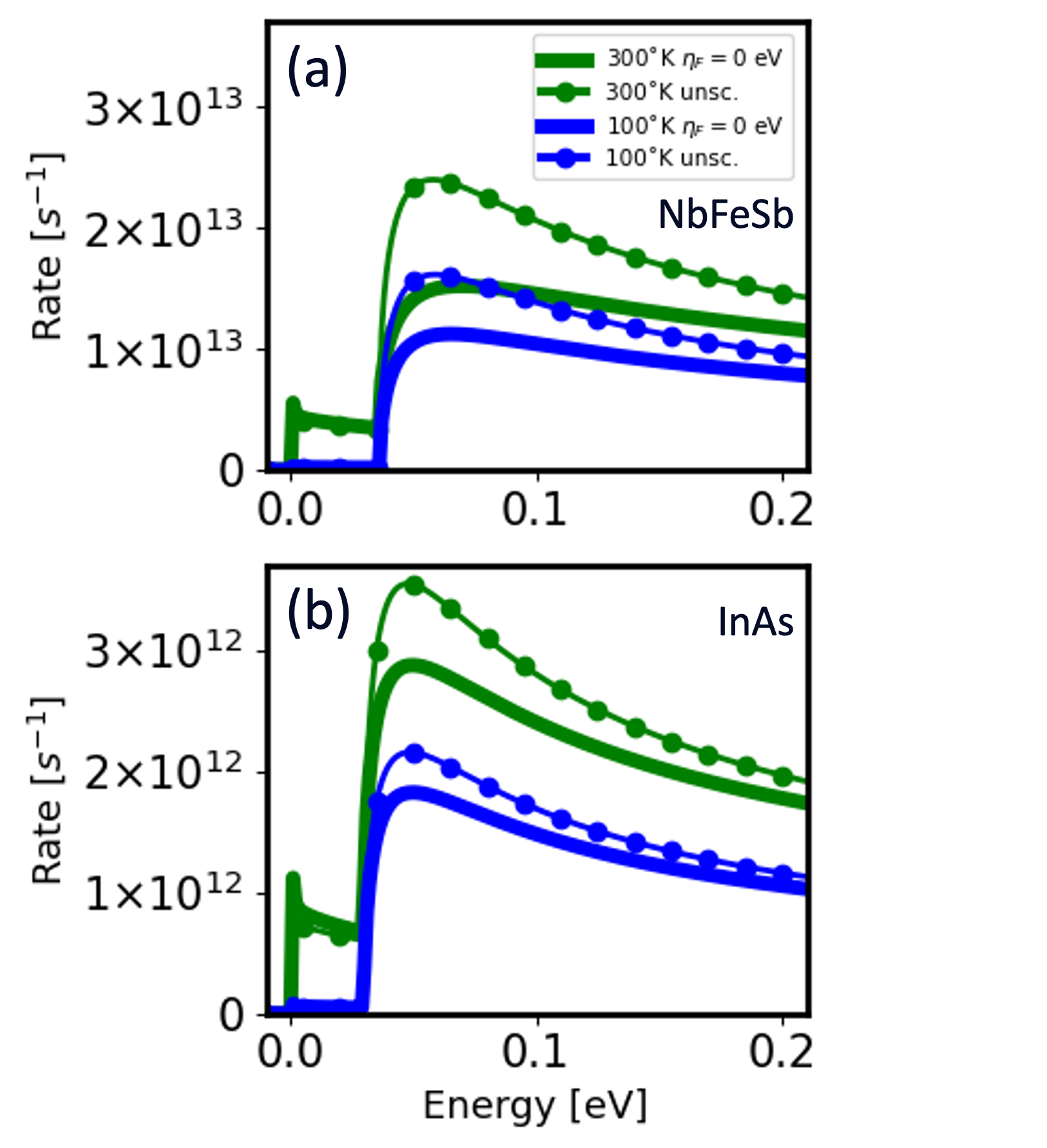}
    \caption{Scattering rates for NbFeSb (a) and InAs (b) at $300^\circ$K (green) and $100^\circ$K (blue). Solid lines: POP with screening ($\eta_F = 0$ eV). Solid-dot line: POP without screening. The scattering rate peak decreases for NbFeSb by $37.8\%$ ($300^\circ$K) and $31.0\%$ ($100^\circ$K), and for InAs by $19.0\%$ ($300^\circ$K) and $15.4\%$ ($100^\circ$K).}
    \label{fig:figS5}
\end{figure}

This is shown in Fig. \ref{fig:figS5}, where the scattering rates for NbFeSb and InAs are compared against each other. The green and blue colors correspond to temperatures of $300^\circ$K and $100^\circ$K respectively, while the solid and solid-dot lines correspond to screened (at $\eta_F = 0$eV) and unscreened POP scattering. The effect of screening is seen to be weaker for InAs compared to NbFeSb (more closely packed dotted and solid lines), especially at $100^\circ$K. 

\bibliography{pop_screen}

\end{document}